%% file: BMDR_Applications_Part_2.tex
\documentclass[12pt, draftclsnofoot, onecolumn]{IEEEtran}
\usepackage[T1]{fontenc}
\usepackage{amssymb}
\usepackage[pdftex]{graphicx}
\usepackage{caption}[2020/09/12]
\usepackage{amsmath}
\usepackage{amsthm}
\usepackage{mathtools}
\usepackage{array}
\usepackage[noadjust]{cite}
\usepackage{algorithm}
\usepackage{algpseudocode}
\usepackage{multirow}
\usepackage{xcolor}
\usepackage[utf8]{inputenc}
\usepackage{textcomp}
\usepackage[hidelinks]{hyperref}
\usepackage[acronym]{glossaries}
\usepackage[binary-units=true]{siunitx}
\algnewcommand{\Initialize}[1]{%
  \State \textbf{Initialize}
  \Statex \hspace*{\algorithmicindent}\parbox[t]{0.96\linewidth}{\raggedright #1}
}

\makeatletter
\let\MYcaption\@makecaption
\makeatother

\algnewcommand{\LineComment}[1]{\(\triangleright\) #1}

\DeclareSIUnit\frame{frame}
\DeclareSIUnit{\kmph}{kmph}
\DeclareSIUnit{\dBm}{dBm}
\DeclareSIUnit{\microsecond}{\micro\second} 
\DeclareSIUnit\PRB{PRB}
\DeclareMathAlphabet{\mathbcal}{OMS}{cmsy}{b}{n}

\DeclareFontFamily{T1}{calligra}{}
\DeclareFontShape{T1}{calligra}{m}{n}{<->s*[1]callig15}{}
\DeclareMathAlphabet\mathcalligra   {T1}{calligra} {m} {n}
\DeclareMathAlphabet\mathzapf       {T1}{pzc} {mb} {it}
\DeclareMathAlphabet\mathchorus     {T1}{qzc} {m} {n}
\DeclareMathAlphabet\mathrsfso      {U}{rsfso}{m}{n}

\input{abbreviations.tex}
\input{macros.tex}

\begin{document}

\title{Bit-Metric Decoding Rate in Multi-User MIMO Systems: Applications}

\author{K. Pavan Srinath and Jakob Hoydis, \textit{Senior Member, IEEE}%

\thanks{K. P. Srinath is with Nokia Bell Labs, 91620 Nozay, France (email: pavan.koteshwar\_srinath@nokia-bell-labs.com), and J. Hoydis is with NVIDIA, 06906 Sophia Antipolis, France (email: jhoydis@nvidia.com). A significant part of this work was done when J. Hoydis was at Nokia Bell Labs, 91620 Nozay, France.}
}

\maketitle

\begin{abstract}
\input{tex/abstract}
\end{abstract}

\glsresetall
\input{tex/introduction}
\input{tex/system_model}
\input{tex/BMDR_CER}

\input{tex/LA}
\input{tex/PLA}
\input{tex/simulation_results}
\input{tex/discussion}

\bibliographystyle{IEEEtran}
\bibliography{IEEEabrv, references}

\end{document}

%% file: abbreviations.tex
\newacronym{3GPP}{3GPP}{3rd Generation Partnership Project}
\newacronym{5GNR}{5G NR}{5G New Radio}

\newacronym{AWGN}{AWGN}{additive white Gaussian noise}
\newacronym{AM}{AM}{arithmetic mean}
\newacronym{AMP}{AMP}{approximate message passing}
\newacronym{ARQ}{ARQ}{automatic repeat request}

\newacronym{BER}{BER}{bit error rate}
\newacronym{BICM}{BICM}{bit-interleaved coded modulation}
\newacronym{BLER}{BLER}{block error rate}
\newacronym{BMDR}{BMDR}{bit-metric decoding rate}
\newacronym{BS}{BS}{base station}

\newacronym{CB}{CB}{code-block}
\newacronym{CDF}{CDF}{cumulative distribution function}
\newacronym{CER}{CER}{codeword error rate}
\newacronym{CESM}{CESM}{capacity ESM}
\newacronym{CNN}{CNN}{convolutional neural network}
\newacronym{Conv2D}{Conv2D}{convolutional}
\newacronym{CSI}{CSI}{channel state information}

\newacronym{DCI}{DCI}{downlink control information}
\newacronym{DMRS}{DMRS}{demodulation reference signal}

\newacronym{eMBB}{eMBB}{enhanced mobile broadband}
\newacronym{EESM}{EESM}{exponential ESM}
\newacronym{ESM}{ESM}{effective SINR metric}

\newacronym{FC}{FC}{fully-connected}

\newacronym{GM}{GM}{geometric mean}

\newacronym{HARQ}{HARQ}{hybrid automatic repeat request}

\newacronym{iid}{i.i.d.\@}{independent and identically distributed}
\newacronym{ISI}{ISI}{inter-symbol interference}

\newacronym{KL}{KL}{Kullback–Leibler}

\newacronym{LA}{LA}{link-adaptation}
\newacronym{LDPC}{LDPC}{low-density parity-check}
\newacronym{LESM}{LESM}{logarithmic ESM}
\newacronym{LMMSE}{LMMSE}{linear minimum mean square error}
\newacronym{LLR}{LLR}{log-likelihood ratio}

\newacronym{MAP}{MAP}{maximum a posteriori}
\newacronym{MCS}{MCS}{modulation and coding scheme}
\newacronym{MI}{MI}{mutual-information}
\newacronym{MIESM}{MIESM}{mutual-information ESM}
\newacronym{MIMO}{MIMO}{multiple-input multiple-output}
\newacronym{ML}{ML}{maximum-likelihood}
\newacronym{MLD}{MLD}{maximum-likelihood detector}
\newacronym{MU-MIMO}{MU-MIMO}{multi-user MIMO}
\newacronym{MSE}{MSE}{mean squared error}

\newacronym{NN}{NN}{neural network}

\newacronym{OFDM}{OFDM}{orthogonal frequency division multiplexing}
\newacronym{OLLA}{OLLA}{outer loop link adaptation}
\newacronym{OLPC}{OLPC}{open-loop-power-control}

\newacronym{PHY}{PHY}{physical layer}
\newacronym{PRACH}{PRACH}{physical random access channel}
\newacronym{PRB}{PRB}{physical resource block}
\newacronym{PUCCH}{PUCCH}{physical uplink control channel}
\newacronym{PUSCH}{PUSCH}{physical uplink shared channel}

\newacronym{QAM}{QAM}{quadrature amplitude modulation}
\newacronym{QPSK}{QPSK}{quadrature phase-shift keying}

\newacronym{RE}{RE}{resource element}
\newacronym{RBIR}{RBIR}{received bit information rate}

\newacronym{SD}{SD}{standard deviation}
\newacronym{SE}{SE}{spectral efficiency}
\newacronym{SGD}{SGD}{stochastic gradient descent}
\newacronym{SIC}{SIC}{soft interference cancellation}
\newacronym{SINR}{SINR}{signal-to-interference-noise ratio}
\newacronym{SISO}{SISO}{single-input single-output}
\newacronym{SLS}{SLS}{system-level simulations}
\newacronym{SNR}{SNR}{signal-to-noise ratio}
\newacronym{SRS}{SRS}{sounding reference signal}
\newacronym{SU-MIMO}{SU-MIMO}{single-user MIMO}
\newacronym{SU-SISO}{SU-SISO}{single-user SISO}

\newacronym{TB}{TB}{transport-block}

\newacronym{wrt}{w.r.t.\@}{with respect to}

\newacronym{ZF}{ZF}{zero-forcing}
\newacronym{RL}{RL}{reinforcement-learning}

\newacronym{UE}{UE}{user-equipment}
\newacronym{uRLLC}{uRLLC}{ultra-reliable low-latency communications}

%% file: macros.tex
\DeclareMathAlphabet{\mathbcal}{OMS}{cmsy}{b}{n}
\renewcommand{\vec}[1]{\mathbf{#1}}
\newcommand{\vecr}[1]{\mathrm{#1}}

\algblock{Input}{EndInput}
\algnotext{EndInput}
\algblock{Output}{EndOutput}
\algnotext{EndOutput}
\newcommand{\Desc}[2]{\State \makebox[3em][l]{#1}#2}

\newcommand{\be}{\begin{equation}}
\newcommand{\ee}{\end{equation}}			
\newcommand{\ben}{\begin{equation*}}
\newcommand{\een}{\end{equation*}}

\newtheorem{definition}{Definition}[section]

\newtheorem{remark}{Remark}

\newtheorem{examp}{Example}[section]

\newcommand{\hv}{\vec{h}}

\newcommand{\nv}{\vec{n}}

\newcommand{\sv}{\vec{s}}

\newcommand{\yv}{\vec{y}}

\newcommand{\Hm}{\vec{H}}
\newcommand{\Id}{\vec{I}}

\newcommand{\Km}{\vec{K}}

\newcommand{\Xm}{\vec{X}}

\newcommand{\Hrm}{\vecr{H}}

\newcommand{\yrv}{\vecr{y}}

\newcommand{\Cc}{{\mathcal{C}}}
\newcommand{\Dc}{{\cal D}}

\newcommand{\Gc}{{\cal G}}
\newcommand{\Hc}{{\cal H}}
\newcommand{\Ic}{{\cal I}}

\newcommand{\Lc}{{\cal L}}
\newcommand{\Mc}{{\cal M}}
\newcommand{\Nc}{{\mathcal{N}}}
\newcommand{\Oc}{{\cal O}}

\newcommand{\Qc}{{\cal Q}}
\newcommand{\Rc}{{\cal R}}
\newcommand{\Sc}{{\cal S}}

\newcommand{\Hbc}{{\mathbcal H}}

\newcommand{\Cbb}{\mathbb{C}}

\newcommand{\LB}{\left(}
\newcommand{\RB}{\right)}
\newcommand{\LP}{\left\{}
\newcommand{\RP}{\right\}}
\newcommand{\LSB}{\left[}
\newcommand{\RSB}{\right]}

\renewcommand{\ln}[1]{\mathop{\mathrm{ln}}\LB #1\RB}
\renewcommand{\log}[1]{\mathop{\mathrm{log}_2}\LB #1\RB}
\newcommand{\logt}[1]{\mathop{\mathrm{log}_{10}}\LB #1\RB}
\renewcommand{\exp}[1]{\mathop{\mathrm{exp}}\LB #1\RB}

\newcommand{\EE}{{\mathbb{E}}}
\newcommand{\Expect}[2]{\EE_{#1}\LSB #2\RSB}
\newcommand{\PP}{{\mathbb{P}}}
\newcommand{\Prob}[1]{\PP \LP #1 \RP}

 \newcommand{\argmax}[1]{\underset{#1}{\operatorname{arg}\,\operatorname{max}}\;}
 \newcommand{\argmin}[1]{\underset{#1}{\operatorname{arg}\,\operatorname{min}}\;}

%% file: tex/abstract.tex
This is the second part of a two-part paper that focuses on \gls{LA} and \gls{PHY} abstraction for \gls{MU-MIMO} systems with non-linear receivers. The first part proposes a new metric, called \gls{BMDR} for a detector, as being the equivalent of post-equalization \gls{SINR} for non-linear receivers. Since this \gls{BMDR} does not have a closed form expression, a machine-learning based approach to estimate it effectively is presented. In this part, the concepts developed in the first part are utilized to develop novel algorithms for \gls{LA}, dynamic detector selection from a list of available detectors, and \gls{PHY} abstraction in \gls{MU-MIMO} systems with arbitrary receivers. Extensive simulation results that substantiate the efficacy of the proposed algorithms are presented.


%% file: tex/introduction.tex
\section{Introduction}
\label{sec:intro}

In the first part~\cite{kps_jh_part1} of this two-part paper, we introduced the notion of \gls{BMDR} for each user in a \gls{MU-MIMO} system with any arbitrary detector. \gls{BMDR} is an information-theoretic measure that takes into consideration the effect of a possibly suboptimal detector. Since \gls{BMDR} does not have a closed form expression that would allow its instantaneous calculation, we described a machine-learning based solution to predict it for an observed set of channel realizations. In the second part of this two-part paper, we use the concepts developed in the first part to address two important problems in \gls{MU-MIMO} systems that are detailed below:

\subsection{Uplink link-adaptation} \label{subsec:Intro_LA}  

\begin{table}[htbp]
    \begin{center}
        \begin{tabular}{|c|c|c|c|c|c|}
        \hline
        \multicolumn{1}{|c|}{\multirow{3}{*}{{\bf Technique}}} & \multicolumn{1}{c|}{\multirow{3}{*}{{\bf Metric}}} &\multicolumn{4}{c|}{{\bf Applicability}} \\  
           \cline{3-6} 
         &  & \multicolumn{2}{c|}{{\bf Linear detectors}} & \multicolumn{2}{c|}{{\bf Non-linear detectors}}  \\
         \cline{3-6}
         & & {\bf SU-MIMO} & {\bf MU-MIMO} &{\bf SU-MIMO} & {\bf MU-MIMO} \\ \hline \hline
         LA with OLLA~\cite{la_2012,la_2015,la_2020} & SINR & Yes & Yes & NA & NA \\  \hline
        $K$-best &  & \multirow{3}{*}{NA} & \multirow{3}{*}{NA} & $K$-best detector only, & $K$-best detector only, \\ 
        adaptive  & BER &  & & only modulation and & only modulation and  \\
        modulation \cite{wenjun15} & & & & no rate adaptation & no rate adaptation  \\  \hline 
        MLD  & Symbol & \multirow{3}{*}{NA} & \multirow{3}{*}{NA} & ML detector only, & \multirow{3}{*}{NA} \\ 
        adaptive  & error & & & only modulation and &   \\
        modulation \cite{shin06} & rate & & & no rate adaptation &  \\  \hline 
         The proposed technique & \gls{BMDR} & Yes & Yes & Yes & Yes \\ \hline
        \end{tabular}
    \end{center}
    \caption{Comparison of the proposed \gls{LA} technique with the state-of-the-art.}
    \label{tab:LA_comparison}
\end{table}

In \gls{5GNR}, when a \gls{UE} wishes to send data to the \gls{BS}, it makes a scheduling request on the \gls{PRACH} for the first time, or on the \gls{PUCCH} if already scheduled. The \gls{BS} then sends an uplink grant in the \gls{DCI} message with information on what \glspl{PRB} to use, and what \gls{MCS} to use for data transmission. The method to determine what \gls{MCS} to use is at the discretion of the \gls{BS}. Following this uplink grant, the \gls{UE} sends data using the mentioned \gls{MCS} on the \gls{PUSCH}. This process is part of \gls{LA}, where the \gls{BS} adapts the \gls{MCS} levels according to the capacity of the link between the \gls{UE} and itself.

Some of the best known works on \gls{LA} are~\cite{la_2012,la_2015,la_2020} for linear receivers, \cite{wenjun15} for the $K$-best detector~\cite{Guo2006}, and \cite{shin06} for the \gls{MLD} (summarized in Table \ref{tab:LA_comparison} where we have also highlighted the metric used by each technique for performing \gls{LA}). All known works in the literature on \gls{LA} for linear detectors propose to perform \gls{OLLA}. \Gls{OLLA} is a technique where the \gls{SINR} is initially estimated, and then a correction to this \gls{SINR} estimate is performed to account for any estimation inaccuracy. This correction is dependent on the number of correctly/incorrectly decoded codewords (through ACK/NACK on the downlink for downlink \gls{LA}). Next, the most appropriate \gls{MCS} level is chosen based on this corrected \gls{SINR}. 

In a \gls{MU-MIMO} system, the principles of \gls{OLLA} can still be used and work well when the users are involved in large file transfers. This is especially true for linear receivers (like the \gls{LMMSE} detector~\cite[Ch. 8]{Tse2005}) that allow the computation of a post-equalization \gls{SINR}. However, the known \gls{MCS} selection techniques that are based on post-equalization \gls{SINR} do not work in systems with non-linear detectors like sphere-decoder~\cite{Viterbo1999, Studer2010} and its fixed-complexity variants~\cite{Barbero2008,Guo2006}. To the best of our knowledge, there is no known technique in the literature to perform \gls{MCS} adaptation for arbitrary non-linear detectors.

\subsection{Physical layer abstraction} \label{subsec:Intro_PLA}  

\begin{figure}
    \centering       
        \includegraphics[scale = 0.55]{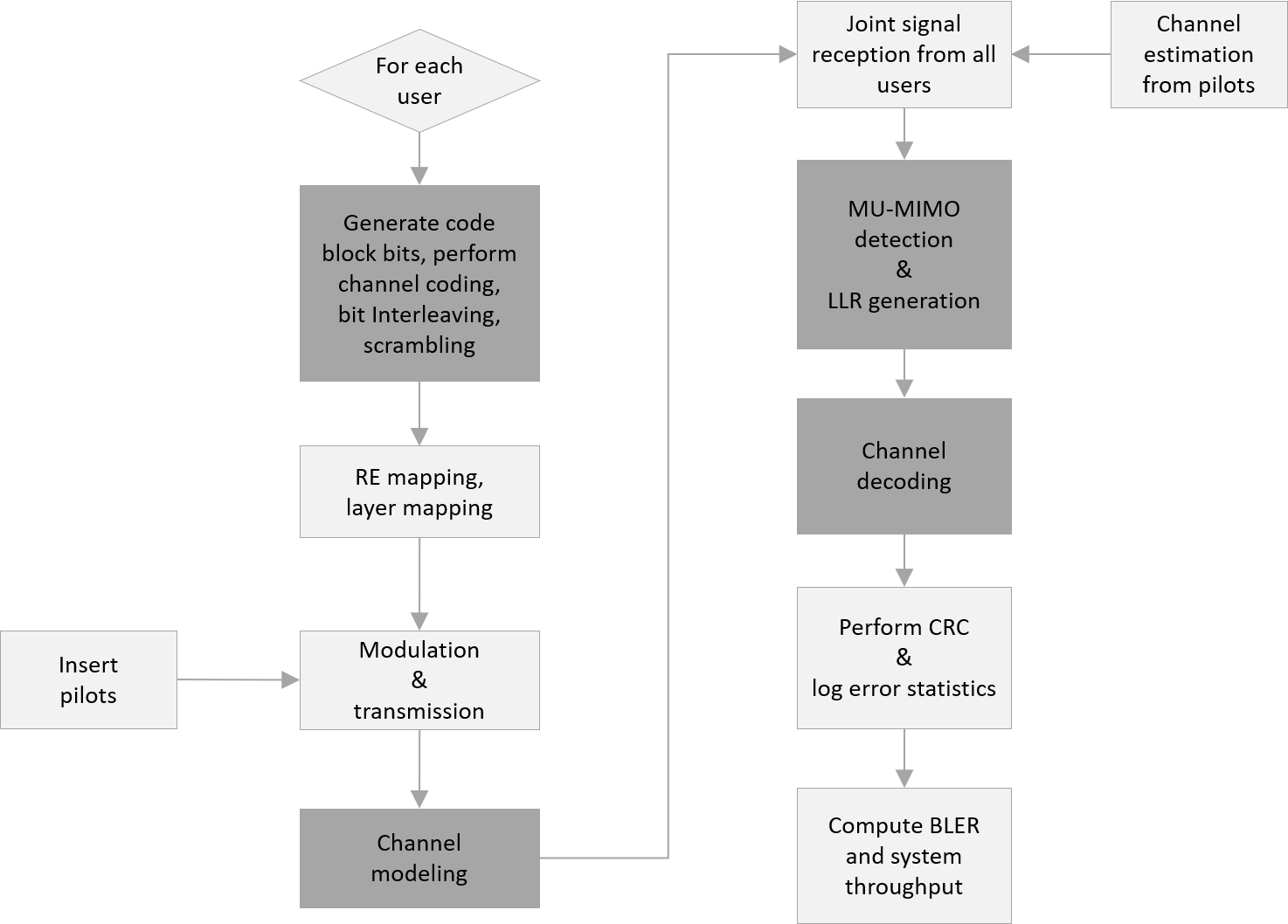}
        \caption{Flowchart of a full MU-MIMO link level simulation. The darkly shaded boxes represent the time-intensive procedures.}
        \label{fig:pla_flowchart_LLS}
\end{figure}

Telecommunication chip manufacturers use \gls{SLS} to evaluate the performance of their algorithms. A typical simulator consists of (but is not limited to) the following functionalities: intercell-interference modeling, resource scheduling and allocation, power allocation and power control, \gls{LA} block with channel quality feedback, channel modeling, link performance modeling. Of these, the link performance-modeling block is the one that models the physical layer components of the communication system. The flowchart of a typical link-level simulation is depicted in Fig.~\ref{fig:pla_flowchart_LLS}. The boxes that are darkly shaded refer to the operations that are time-intensive/resource-intensive to execute, and these include (but are not limited to) \gls{CB} segmentation, channel-coding, bit-interleaving, and scrambling at the transmitter side, and \gls{MIMO} detection with \gls{LLR} generation, and channel decoding at the receiver side.  Therefore, in order to reduce the complexity of \gls{SLS}, these components are replaced by simpler functionalities that are quicker to execute but capture the essential behavior of the overall physical layer. This technique is called \gls{PHY} abstraction. To be precise, the goal of link performance modeling (or equivalently, \gls{PHY} abstraction) is to obtain the same figures of merit for performance evaluation as would be obtained if the original components were used, but with much simpler complexity. The commonly used figures of merit are system throughput and \gls{CER} or \gls{BLER}, noting that a block can consist of multiple codewords~\cite[Section 5]{3GPP_coding_2020}.

\begin{table}[htbp]
    \begin{center}
        \begin{tabular}{|c|c|c|c|c|c|}
        \hline
        \multicolumn{1}{|c|}{\multirow{3}{*}{{\bf Technique}}} & \multicolumn{1}{c|}{\multirow{3}{*}{{\bf Metric}}} &\multicolumn{4}{c|}{{\bf Applicability}} \\  
           \cline{3-6} 
         &  & \multicolumn{2}{c|}{{\bf Linear detectors}} & \multicolumn{2}{c|}{{\bf Non-linear detectors}}  \\
         \cline{3-6}
         & & {\bf SU-MIMO} & {\bf MU-MIMO} &{\bf SU-MIMO} & {\bf MU-MIMO} \\ \hline \hline
         EESM, MIESM/RBIR~\cite{phy_abs_2005,phy_abs_2006,lagen2021new} & \gls{SINR} & Yes & Yes & NA & NA \\  \hline
          &  & Yes, & Yes, &  &  \\
          The proposed technique & BMDR & equivalent to & equivalent to & Yes &  Yes \\
         & & MIESM/RBIR & MIESM/RBIR & & \\ \hline
        \end{tabular}
    \end{center}
    \caption{Comparison of the proposed \gls{PHY} abstraction technique with the state-of-the-art.}
    \label{tab:Phy_abstraction_comparison}
\end{table}

In the literature, there exist several abstraction models for the case of \gls{SU-SISO} systems, and (with some limitations) for \gls{MIMO} systems with linear receivers. A few papers on this topic are~\cite{phy_abs_2005,phy_abs_2006,lagen2021new} and references therein (summarized in Table \ref{tab:Phy_abstraction_comparison}). The limitations of the approaches in these papers are the following:
\begin{itemize}    
    \item For linear detectors, these papers propose to compress the set of post-equalization \glspl{SINR} obtained at the receiver over every \gls{RE} into an effective \gls{SINR}, and then map this effective \gls{SINR} to a \gls{BLER} using an approximate \gls{SINR}-\gls{BLER} lookup table. This compression of several post-equalization \glspl{SINR} to a single effective \gls{SINR} is known as \gls{ESM}. There are several such metrics in the literature: \gls{EESM}, \gls{MIESM}, \gls{CESM}, and \gls{LESM}. A variant of \gls{MIESM}, known as \gls{RBIR}, is proposed for usage in IEEE 802.11 while 3GPP recommends \gls{EESM}. However, there is no clear consensus on the selection of a single method. 
    \item For \gls{MIMO} systems with non-linear detectors, there is no known technique in the literature to perform \gls{PHY} abstraction. 
\end{itemize}

\noindent These limitations are addressed in this part of the two-part paper. The summary of our contributions is listed below:
\begin{itemize}
\item We describe a new algorithm (Section~\ref{sec:la}) for performing \gls{LA} in \gls{MU-MIMO} systems for arbitrary detectors. This algorithm makes use of \gls{BMDR} in lieu of post-equalization \gls{SINR}. 
\item We present a technique to dynamically select the most appropriate detector from a list of available detectors (Section~\ref{subsec:detector_selection}).
\item We propose a new method (Section~\ref{sec:pla}) for performing \gls{PHY} abstraction in \gls{MU-MIMO} systems with arbitrary receivers. 
\item Extensive simulation results are provided to verify the efficacy of the proposed techniques.  
\end{itemize} 

\subsection*{Paper Organization}
The system model and a few relevant definitions are presented in Section~\ref{sec:system_model}. Section~\ref{sec:bmdr} explains the technique to obtain a \gls{BMDR}-\gls{CER} map. Section~\ref{sec:la} presents a new algorithm for \gls{LA} and dynamic detector selection while Section~\ref{sec:pla} describes a new technique to perform \gls{PHY} abstraction. Simulation results showing the efficacy of the proposed techniques are presented in Section~\ref{sec:sim_results}, and concluding remarks constitute Section~\ref{sec:conc_remarks}. 

\subsection*{Notation}
Boldface upper-case (lower-case) letters denote random matrices (vectors), and normal upright upper-case (lower-case) letters are understood from context to denote the realizations of random matrices (vectors). The field of complex numbers is denoted by $\Cbb$. The notation $\Xm \in \Sc^{m \times n}$ denotes that $\Xm$ is a matrix of size $m \times n$ with each entry taking values from a set $\Sc$, and $\Xm^{\mathrm{T}}$ represents the transpose of $\Xm$. The identity matrix of size $n\times n$ is denoted by $\Id_n$, and $\nv \sim \Cc\Nc( 0, \Id_n)$ indicates that $\nv$ is sampled from the $n$-dimensional complex standard normal distribution.

%% file: tex/system_model.tex
\section{System Model and Definitions}
\label{sec:system_model}

We consider a \gls{BICM} system \cite{Caire1998} for an \gls{OFDM} based \gls{MU-MIMO} uplink transmission as in the first part \cite[Section~II]{kps_jh_part1}, and briefly reiterate the setup here. There are $N_u$ \glspl{UE} that transmit data to a \gls{BS} on the same set of resources. The $i^{th}$ \gls{UE} uses a channel code with code-rate $r_i$ and codeword length $n_i$. There are $n_t^{(i)}$ transmit antennas at \gls{UE} $i$ which uses a constellation $\Qc_i$ of cardinality $2^{m_i}$ for some $m_i \in \LP 1, 2, 4, 6, \cdots \RP$ so that groups of $m_i$ codeword bits are mapped to a constellation symbol. With $\sum_{i=1}^{N_u}n_t^{(i)}=N$, the signal model is given as 
\begin{equation}\label{eq:signal_model}
\yv_{f,t} = \sum_{i=1}^{N_u}\sqrt{\frac{\rho_i}{n_t^{(i)}}}\Hm_{f,t,i}\sv_{f,t,i} + \nv_{f,t} = \Hm_{f,t}\sv_{f,t} + \nv_{f,t} 
\end{equation}
where $\yv_{f,t} \in \Cbb^{n_r \times 1}$ is the received signal vector at the \gls{BS} equipped with $n_r$ receive antennas, $\Hm_{f,t,i} \in \Cbb^{n_r \times N}$ is the \gls{MIMO} channel from UE $i$ to the \gls{BS},   $\rho_i$ is the total transmit power of \gls{UE} $i$, $\sv_{f,t,i}\in \Qc_i^{n_t^{(i)} \times 1}$ is the transmitted signal vector from \gls{UE} $i$ with each entry taking values from $\Qc_i$, $\Hm_{f,t} \triangleq \LSB \sqrt{{\rho_1}/{n_t^{(1)}}}\Hm_{f,t,1}, \cdots, \sqrt{{\rho_{N_u}}/{n_t^{(N_u)}}}\Hm_{f,t,N_u} \RSB \in \Cbb^{n_r \times N}$ is the composite \gls{MU-MIMO} channel, $\sv_{f,t} \triangleq [\sv_{f,t,1}^{\mathrm{T}}, \cdots, \sv_{f,t,N_u}^{\mathrm{T}} ]^{\mathrm{T}}$ $\in \Qc^{N \times 1}$ is the composite transmitted signal vector with $ \Qc^{N \times 1} \triangleq \Qc_1^{n_t^{(1)} \times 1} \times \cdots \times \Qc_{N_u}^{n_t^{(N_u)} \times 1}$, and $\nv_{f,t} \sim \Cc\Nc\LB 0, \Id_{n_r}\RB$ represents the complex \gls{AWGN}. The subscript $(f,t)$ denotes the (subcarrier, time) indices of the \gls{RE} in which $\yv_{f,t}$ is received. In each \gls{RE}, \gls{UE} $i$ transmits a total of $m_in_t^{(i)}$ codeword bits. The system model can be better illustrated using the following example setup.

\begin{examp}
    Consider a \gls{MU-MIMO} system with $n_r = 64$ receive antennas at the \gls{BS} and two transmitting \glspl{UE} ($N_u = 2$) that transmit on the same set of resources consisting of $120$ subcarriers and $14$ time symbols. Let \gls{UE} $1$ use constellation $\Qc_1$ to be QPSK ($m_1 = 2$) and \gls{UE} $2$ use constellation $\Qc_2$ to be $16$-QAM ($m_2=4$). Suppose that \gls{UE} $1$ uses one transmit antenna ($n_t^{(1)}=1$) and \gls{UE} $2$ uses three transmit antennas ($n_t^{(2)}=3$) so that $N = 4$. If the channel between \gls{UE} $1$ and the \gls{BS} on \gls{RE} $(f,t)$ is denoted by $\hv_{f,t,1} \in \Cbb^{64 \times 1}$ and that between \gls{UE} $2$ and the \gls{BS} by $\Hm_{f,t,2} \in \Cbb^{64 \times 3}$, the system model is
    \begin{align}
        \yv_{f,t} & = \sqrt{\rho_1}\hv_{f,t,1}s_{f,t,1} + \sqrt{\frac{\rho_2}{3}}\Hm_{f,t,2}\sv_{f,t,2} +  \nv_{f,t} = \Hm_{f,t}\sv_{f,t} + \nv_{f,t},  
    \end{align}
    $f\in \{1.\cdots, 120\}, t\in\{1\cdots, 14\}$, where $\Hm_{f,t} \triangleq \LSB \sqrt{\rho_1}\hv_{f,t,1}, \sqrt{{\rho_{2}}/{3}}\Hm_{f,t,2} \RSB \in \Cbb^{64 \times 4}$, $s_{f,t,1} \in \Qc_1$, $\sv_{f,t,2} \in \Qc_2^{3\times1}$, and $\sv_{f,t} \triangleq \LSB s_{f,t,1}, \sv_{f,t,2}^{\mathrm{T}} \RSB^{\mathrm{T}}$. In particular, $\sv_{f,t}$ is of the form $[s_a, s_b, s_c, s_d]^{\mathrm{T}}$ where $s_a$ is a QPSK symbol while $s_b,s_c,$ and $s_d$ are symbols from $16$-QAM. In each \gls{RE}, \gls{UE} $1$ transmits $m_1n_t^{(1)}= 2$ bits of coded data while \gls{UE} $2$ transmits $m_2n_t^{(2)} = 12$ bits of coded data.
\end{examp}

In practice, there will be intercell-interference (from neighboring cell users) and imperfect channel estimation at the serving \gls{BS}. Let $\hat{\Hm}_{f,t}$ denote the estimated channel so that $\Hm_{f,t} = \hat{\Hm}_{f,t} + \Delta{\Hm}_{f,t}$ where $\Delta{\Hm}_{f,t}$ denotes the estimation error. The signal model of~\eqref{eq:signal_model} in the presence of interference noise from neighboring cells can be written as
\begin{align}\label{eq:imperf_est1}
\yv_{f,t} = \hat{\Hm}_{f,t}\sv_{f,t} + \Delta{\Hm}_{f,t}\sv_{f,t} + \nv_{f,t}
\end{align} 
where the interference noise is subsumed in $\nv_{f,t}$ so that $\nv_{f,t} \sim \Cc\Nc\LB 0, \Km_n \RB$ with $\Km_n \in \Cbb^{n_r \times n_r}$ being a (known) Hermitian, positive-definite but non-diagonal matrix. Assuming that an \gls{LMMSE} estimator is used for channel estimation with a known estimation-error covariance of $\Km_e$, we arrive at the following signal model (refer \cite[Section~II-A]{kps_jh_part1} for details) after noise-whitening: 
\begin{align}\label{eq:imperf_est2}
\yv'_{f,t} = {\Hm'}_{f,t}\sv_{f,t} + \nv_{f,t}' 
\end{align} 
where $\yv'_{f,t} = \LB \Km_n + \Km_e \RB^{-\frac{1}{2}}\yv_{f,t}$, ${\Hm'}_{f,t} = \LB \Km_n + \Km_e \RB^{-\frac{1}{2}}\hat{\Hm}_{f,t}$, and $\nv'_{f,t} \sim \Cc\Nc\LB 0, \Id_{n_r}\RB$ due to the noise-whitening. Therefore, unless specified otherwise, the assumed signal model is that of~\eqref{eq:signal_model}, and any other realistic model can be converted to this form. 

Let $b_{f,t,i,l,j}$ denote the $j^{th}$ transmitted bit of \gls{UE} $i$ on its $l^{th}$ transmit antenna (a detailed system model is available in \cite[Section~II]{kps_jh_part1}). Suppose that a \gls{MU-MIMO} detector $\Dc$ is used to generate the \gls{LLR} for each $b_{f,t,i,l,j}$, $i = 1,\cdots,N_u$, $j=1,\cdots, m_i$, $l=1,\cdots,n_t^{(i)}$. Let $q_{\Dc}(b_{f,t,i,l,j}; \yrv_{f,t}, \Hrm_{f,t})$ denote the posterior probability $\Prob{b_{f,t,i,l,j} \vert \yv_{f,t}, \Hm_{f,t}, \Dc }$ obtained from the LLR for $b_{f,t,i,l,j}$. 

\begin{definition}
    {\it \gls{BMDR} \cite[Section~III]{kps_jh_part1}}:  The \gls{BMDR} of a detector $\Dc$ for \gls{UE} $i$ for a channel matrix $\Hm_{f,t}$, denoted by $R_{\Dc,i}(\Hm_{f,t})$, is defined to be
    \begin{eqnarray} \label{eq:gen_bmdr_channel_def} 
        R_{\Dc,i}(\Hm_{f,t})  & \triangleq  & \max\LP 1 +\Expect{\yv_{f,t} \vert \Hm_{f,t} }{\frac{1}{m_in_t^{(i)}}\sum_{l=1}^{n_t^{(i)}}\sum_{j=1}^{m_i} \log{q_{\Dc}(\boldsymbol{b}_{f,t,i,l,j}; \yv_{f,t}, \Hm_{f,t})}}, 0\RP
       \end{eqnarray}
    where $\LP \boldsymbol{b}_{f,t,i,l,j} \RP_{l,j}$ are related to the elements of $\sv_{f,t,i}$ through a bijective map that assigns groups of $m_i$ bits to a symbol in $\Qc_i$, and $\yv_{f,t}$ is dependent only on $\sv_{f,t}$ and $\nv_{f,t}$ when conditioned on $\Hm_{f,t}$. Note that $R_{\Dc,i}(\Hm_{f,t})$ is itself a random variable whose realization is dependent on the realization of the random channel matrix $\Hm_{f,t}$. Similarly, the BMDR of a detector $\Dc$ for a set of channel matrices $\Hbc$ is defined as
    \begin{eqnarray} \label{eq:gen_bmdr_def} 
        R_{\Dc,i}(\Hc)  & \triangleq  &  \frac{1}{\vert \Hc \vert}\sum_{\Hm_{f,t} \in \Hc}  R_{\Dc,i}(\Hm_{f,t}).
       \end{eqnarray} 
\end{definition}

 \gls{BMDR} as given by \eqref{eq:gen_bmdr_def} represents the normalized (by the codeword length) mutual information between the transmitted bits and the detector output conditioned on a set of channel realizations  $\Hc$ \cite[Theorem 1]{kps_jh_part1} for a \gls{BICM} system. The significance of this is that if the codeword bits of \gls{UE} $i$ are transmitted over a certain set of channel realizations $\Hc$ and detector $\Dc$ is used to generate the \glspl{LLR}, the probability of \gls{CER} can be made arbitrarily close to $0$ only if the rate $r_i$ of the channel code is below $R_{\Dc,i}(\Hc)$.  

Suppose that the codeword bits of \gls{UE} $i$ are transmitted over a set of channel realizations $\Hc$. In the rest of the paper, we focus only on \gls{QAM} due to its significance in most wireless communications standards; so, the modulation order $m$ specifies $2^m$-QAM. Let $\mathfrak{m} \triangleq \LP m_1,\cdots,m_{N_u} \RP$ be the set of modulation orders used by all the co-scheduled users. To emphasize that the \gls{BMDR} depends on $\mathfrak{m}$, we use a slight change of notation from \eqref{eq:gen_bmdr_def}, and denote the \gls{BMDR} of \gls{UE} $i$ for an observed channel realization $\Hm_{f,t}$ by $R_{\Dc,i}\LB \mathfrak{m}, \Hm_{f,t}\RB$, and that for $\Hc$ by $R_{\Dc,i}\LB \mathfrak{m}, \Hc\RB = \frac{1}{\vert \Hc \vert}\sum_{\Hm_{f,t} \in \Hc}R_{\Dc,i}\LB \mathfrak{m}, \Hm_{f,t}\RB$. The result of \cite[Theorem~1]{kps_jh_part1} is that in order to achieve a low \gls{CER}, it is necessary that $R_{\Dc,i}\LB \mathfrak{m}, \Hc\RB > r_i$. Even though $R_{\Dc,i}\LB \mathfrak{m}, \Hc\RB$ cannot be known in advance, its value can be predicted from the most recent observations of the channel. This is the main idea behind the usage of \gls{BMDR} for \gls{LA}.

%% file: tex/BMDR_CER.tex
\section{Mapping BMDR to CER}
\label{sec:bmdr}

\begin{figure}
    \centering       
        \includegraphics[scale = 0.5]{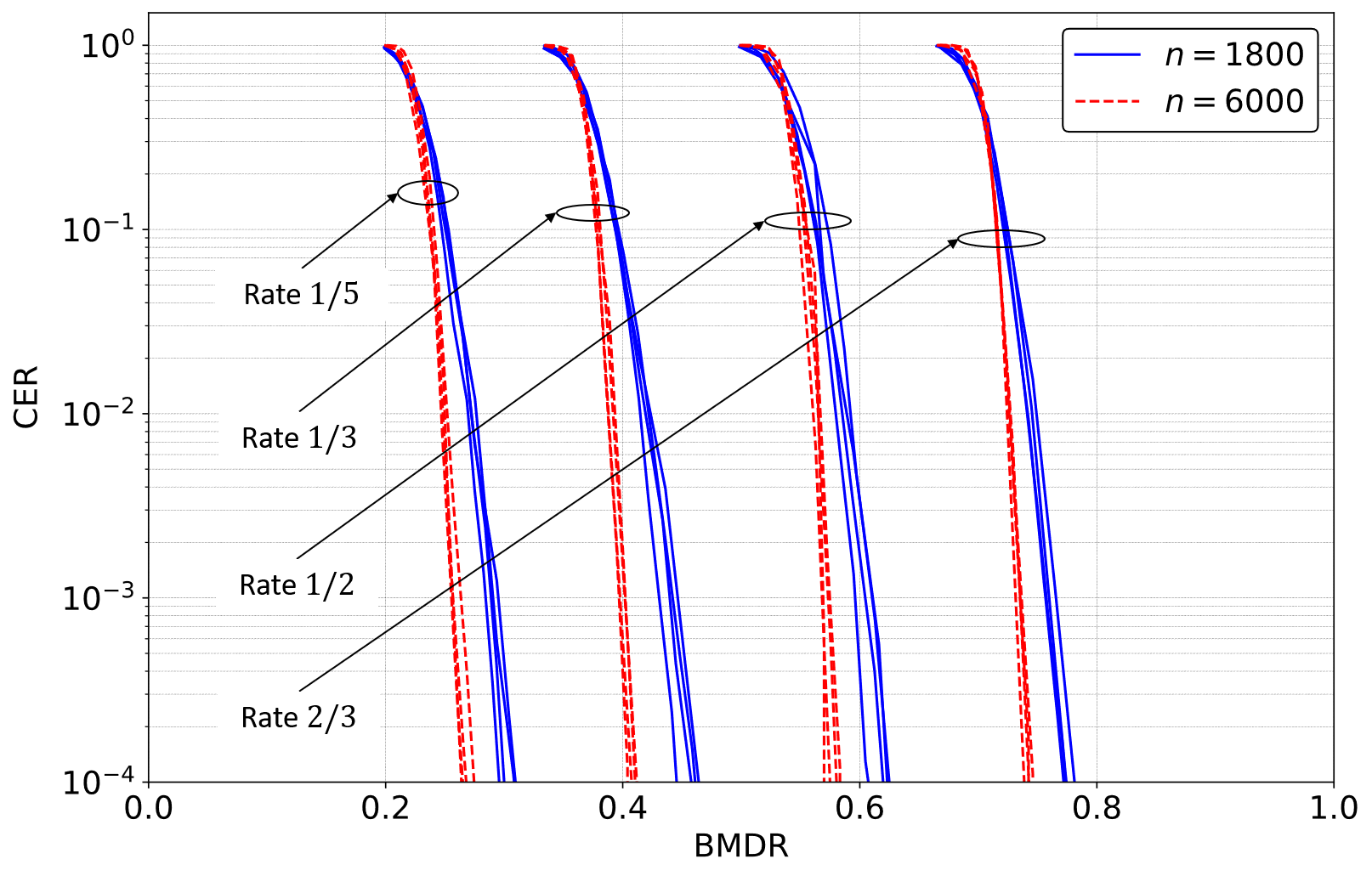}
        \caption{\gls{BMDR} vs \gls{CER} for the \gls{SISO}-\gls{AWGN} channel using $4/16/64/256-$QAM and different code-rates. The solid blue lines correspond to a length-$1800$ 5G-LDPC code while the dashed red lines correspond to a length-$6000$ 5G-LDPC code.}
        \label{fig:bmdr_cer_awgn}
\end{figure}

 Before detailing the algorithm for \gls{LA}, it is necessary to describe a method to map \gls{BMDR} to \gls{CER}. Fig.~\ref{fig:bmdr_cer_awgn} shows the plots of the \gls{CER} as a function of \gls{BMDR} for the \gls{SISO}-\gls{AWGN} channel with the \gls{MLD}, for length-$1800$ and length-$6000$ 5G-\gls{LDPC} codes of various rates. The plots were numerically obtained for $4/16/64/256-$QAM (with the smaller constellation slightly better than the larger constellation). For each modulation order, we chose a range of \gls{SNR} values, and computed the \gls{CER} of a particular length-$n$, rate-$r$ \gls{LDPC} code for a set of uniformly spaced \gls{SNR} values within the chosen range. We then empirically computed the \gls{BMDR} for the same modulation order and for the same set of chosen \gls{SNR} values. After numerically obtaining the \gls{BMDR}-\gls{CER} pair for each \gls{SNR} value, we plotted CER as function of \gls{BMDR} for each code and modulation order. {\it It is to be noted that since we are using the \gls{MLD}, for any \gls{SNR} value, the corresponding \gls{BMDR} multiplied by $m$ is the capacity of \gls{BICM} for $2^m$-QAM~\cite{Caire1998}} at that \gls{SNR}. Therefore, even though the \gls{BMDR} curves for the same value of $n$ and $r$ and for different modulation orders nearly coincide, the overall capacity of the system is $m$ times the \gls{BMDR} for any particular modulation order $m$. To illustrate further with an example, let us consider the rate-$1/3$ channel coding scheme with codeword length $n=1800$ and modulation orders $2$ and $4$ (corresponding to QPSK and $16$-QAM). For a target \gls{CER} of $10^{-3}$, the corresponding \gls{BMDR} for both the modulation orders is approximately $0.45$. This means that while a \gls{CER} of $10^{-3}$ was observed for QPSK at a certain \gls{SNR} $\rho_a$ and that for $16$-QAM at a different \gls{SNR} $\rho_b$ (with $\rho_b > \rho_a$), the observed \gls{BMDR} with QPSK at an \gls{SNR} of $\rho_a$ was the same as the observed \gls{BMDR} with $16$-QAM at an \gls{SNR} of $\rho_b$. It is also to be noted that with $16$-QAM, the number of transmitted coded bits per channel use is twice that with QPSK, albeit at a higher \gls{SNR}, in order to achieve the same \gls{CER}. The other important observation from the plots is that the same target \gls{CER} can be achieved with lower \gls{BMDR} by increasing the codeword length $n$. These observations suggest that for a given code-rate, \gls{BMDR} is strongly dependent on the codeword length and weakly dependent on the modulation order.

We propose to generate a lookup table using simulations performed on a \gls{SISO} \gls{AWGN} channel. We expect this lookup table to be relevant even to the case of \gls{MU-MIMO} because the proof of \cite[Theorem~1]{kps_jh_part1} suggests that the error behavior of a coding scheme depends predominantly on \gls{BMDR} alone. In the rest of the paper, to distinguish between codes from the same family (like the 5G-\gls{LDPC} codes), we explicitly state the rate and length; so $\Cc(r, n)$ denotes a code with rate $r$ and length  $n$. Let $\mathzapf{R}\LB m, \Cc(r, n), \epsilon \RB$ denote the target \gls{BMDR} required to be guaranteed of a \gls{CER} of at most $\epsilon$ for the \gls{SISO}-\gls{AWGN} channel using $2^m$-QAM, $\Cc(r, n)$, and the \gls{MLD}. With this, for a low target probability of codeword error $\epsilon$ ($< 10^{-1}$), we can expect the following.
\begin{enumerate}
    \item $\mathzapf{R}\LB m, \Cc(r, n_1), \epsilon \RB \leq \mathzapf{R}\LB m, \Cc(r, n_2), \epsilon \RB$ if $n_1 > n_2$.   
    \item $\mathzapf{R}\LB m_1, \Cc(r, n), \epsilon \RB \approx \mathzapf{R}\LB m_2, \Cc(r, n) , \epsilon \RB$ if $ m_1 \neq  m_2 $. 
\end{enumerate}

\begin{algorithm} 
    \caption{Pseudocode to generate a look-up table that maps \gls{AWGN} \gls{BMDR} to \gls{CER}.}
    \label{alg:lookup_table}
    \begin{algorithmic}
      \Input      
      \Desc{$\Cc(r, n)$}{: Channel code with code-rate $r$ and length $n$}  
      \Desc{$m$}{: Modulation order used, implies $2^m$-QAM}  
      \Desc{$\mathfrak{P}$}{: Discrete set of usable \gls{SNR} values in \si{\dB}, \\ ~~~~~~~~~~~~~~ $\mathfrak{P}= \LP \rho_{min} + l \Delta\rho, ~l = 0,1,\cdots, \left\lfloor\frac{\rho_{max} - \rho_{min}}{\Delta \rho} \right\rfloor \RP$}    
      \EndInput
      \Output
      \Desc{$\Lc_{U}$}{: Look-up table containing \gls{BMDR}-\gls{CER} pairs}      
      \EndOutput     
    \Initialize{\strut$\Lc_{U} \gets \LP \RP$ } 
    \ForAll{$\rho \in \mathfrak{P}$}
        \State /* \textit{Empirically compute CER for SISO-AWGN channel with channel code $\Cc(r, n)$, \\~~~~~~~  $2^m$-QAM, and maximum-likelihood detection at an SNR of $\rho$ \si{\dB} } */
        \State $P_e \gets \mathrm{CER}\Big($SISO-AWGN, $2^m$-QAM, $\Cc(r,n)$, $\rho \Big)$
        \State /* \textit{Empirically compute BMDR for SISO-AWGN channel with $2^m$-QAM and \\~~~~~~ maximum-likelihood detection at an SNR of $\rho$ \si{\dB} } */
        \State $R_{MLD} \gets \mathrm{BMDR}\Big($SISO-AWGN, $2^m$-QAM, $\rho\Big)$
        \State $\Lc_{U} \gets \Lc_{U}\cup \LP \LB R_{MLD}, P_e\RB \RP$
    \EndFor    
    \end{algorithmic}
\end{algorithm}

 Algorithm~\ref{alg:lookup_table} provides pseudocode for the lookup table generation technique. In the algorithm, \gls{BMDR} is empirically calculated using the method outlined in \cite[Algorithm 1]{kps_jh_part1}, but for a \gls{SISO}-\gls{AWGN} channel. Since the effect of QAM size on the target \gls{BMDR} appears to be very small, one can use just QPSK in Algorithm~\ref{alg:lookup_table} instead of higher dimensional $2^m$-QAM in order to reduce the computational complexity. Having generated the lookup table $\Lc_U$ as detailed in Algorithm~\ref{alg:lookup_table}, $\mathzapf{R}\LB m, \Cc(r, n), \epsilon \RB$ for a target \gls{CER} of $\epsilon$ can be obtained from $\Lc_U$ as 
\begin{equation}\label{eq:target_bmdr}
    \mathzapf{R}\LB m,\Cc(r, n), \epsilon \RB = \min_{\LB R_{MLD}, P_e \RB \in \Lc_U}\LP R_{MLD}~ \vert ~ P_e \leq \epsilon \RP.  
\end{equation}

It is common to have several choices for the code-length $n$ for a given code-rate $r$. For example, in \gls{5GNR}, $k \in \{1,8448\}$~\cite[Section 5]{3GPP_coding_2020} so that $1/r \leq n \leq 8448/r$. It would be practically infeasible to generate a lookup table for each combination of $n$ and $r$. Instead, one can compute $\mathzapf{R}\LB m, \Cc(r, n), \epsilon \RB$ for a few values of $n \in [n_{min}, n_{max}]$ (extremes included), where $n_{max}$ and $n_{min}$ denote the maximum and minimum code-lengths, respectively. From these computed values, any target \gls{BMDR} for intermediate lengths ($n_{min} < n < n_{max}$) can be estimated by linear or polynomial interpolation. We do not go into the details of this and assume that for any target \gls{CER} $\epsilon$, code-rate $r$, modulation order $m$, and code-length $n_{min} \leq n \leq n_{max}$, there exists a (possibly learned) function $\mathzapf{R}$ which outputs the estimated target \gls{BMDR} for $\Cc(r,n)$ with $2^m$-QAM.

%% file: tex/LA.tex
\section{Link-Adaptation in \gls{MU-MIMO} Systems}
\label{sec:la}

Let $\Mc$ denote the set of available modulation orders, and $\Rc(m)$ denote the set of available code-rates for $m\in \Mc$. We also assume for the sake of simplicity that the codewords of all the users are to be transmitted over the same set of \glspl{RE} with the corresponding set of channel realizations denoted by $\Hc_{next}$. This means that if \gls{UE} $i$ selects a modulation order $m_i$, the code-length of the channel code that it uses is $\vert \Hc_{next} \vert m_in_t^{(i)}$. Assuming the usage of an arbitrary detector $\Dc$ and a target \gls{CER} of $\underline{\epsilon}$ for all users, \gls{MCS} selection aims to solve the following joint \gls{SE}-maximizing problem.
\begin{equation}\label{eq:opt_MCS}
  \LP \LB m_i^*, r_i^* \RB\RP_{i=1}^{N_u} = \argmax{\LP m_i \in \Mc, r_i \in \Rc(m_i)\RP_{i=1}^{N_u}}\LP \frac{1}{N_u}\sum_{i=1}^{N_u}r_i m_i~ \Big \vert ~ R_{\Dc,i}\LB \mathfrak{m},\Hc_{next}\RB \geq \mathzapf{R} \LB m_i, \Cc(r_i, n_i), \underline{\epsilon} \RB  \RP
\end{equation}
where $\mathfrak{m} = \LP m_1,\cdots,m_{N_u} \RP$.	However, we do not yet have the estimates of the channel matrices of $\Hc_{next}$, and can only rely on the most recent past-channel estimates. In \gls{5GNR}, these can be the channel estimates obtained using \glspl{DMRS} in the previous transmission slot for those users that have already transmitted, or using \glspl{SRS} if there was no transmission in the previous slot. We denote the set of these channel estimates by $\widehat{\Hc}$. Using a trained \gls{BMDR}-predictor \cite[Section V]{kps_jh_part1}, the \gls{BMDR} for \gls{UE} $i$ can be predicted for $\widehat{\Hc}$ as $\hat{R}_{\Dc,i}\LB \mathfrak{m},\widehat{\Hc} \RB$. However, since these are estimates on a possibly outdated set of channel estimates, we need a \gls{BMDR}-correction offset $\delta_i$ for each \gls{UE} $i$. This correction offset is similar to the \gls{SINR}-correction offset in \gls{OLLA}, and it captures the amount of confidence one has in the estimated \gls{BMDR}; the greater the confidence, the closer $\delta_i$ is to $0$. In case the \gls{BMDR} is underestimated, a positive $\delta_i$ acts as a correction. In large file transfers, $\delta_i$ typically varies throughout the transmission; decreasing by a large value for every incorrectly decoded codeword, and increasing by a small value otherwise. With this, the practical \gls{MCS}-selection problem can be restated as
\begin{equation}\label{eq:practical_opt_MCS}
  \LP \LB m_i^*, r_i^* \RB\RP_{i=1}^{N_u} = \argmax{\LP m_i \in \Mc, r_i \in \Rc(m_i)\RP_{i=1}^{N_u}}\LP \frac{1}{N_u}\sum_{i=1}^{N_u}r_i m_i~ \Big \vert ~ \hat{R}_{\Dc,i}\LB \mathfrak{m},\widehat{\Hc} \RB \geq \mathzapf{R} \LB m_i, \Cc(r_i, n_i), \underline{\epsilon} \RB - \delta_i  \RP.
\end{equation}
In general, the worst case complexity of an algorithm that solves \eqref{eq:practical_opt_MCS} using brute-force search is $\Oc\LB n_{MCS}^{N_u}\RB$, where $n_{MCS}$ is the number of available \gls{MCS} levels. But, if the detector $\Dc$ were such that for any $i \in \LP 1, \cdots, N_u \RP$, 
\begin{align}\label{eq:detector_separable}
   \hat{R}_{\Dc,j}\LB \mathfrak{m},\Hrm \RB = \hat{R}_{\Dc,j}\LB \mathfrak{m}', \Hrm \RB, \forall j \neq i
\end{align}
where $\mathfrak{m}$ and $\mathfrak{m}'$ only differ in modulation order for \gls{UE} $i$. In other words, if the detector is such that when the constellation of one \gls{UE} is changed and those of the others are kept the same, only the \gls{BMDR} of that particular \gls{UE} changes while those of the other users are unaffected, it is possible to solve \eqref{eq:practical_opt_MCS} efficiently with much lower complexity. Algorithm~\ref{alg:la} provides pseudocode to do so for such detectors. In the algorithm, when calculating $r_i^*$ for a chosen $m_i^*$, if there is no $r_i \in \Rc(m_i^*)$ satisfying $\hat{R}_{\Dc,i}\LB \mathfrak{m},\widehat{\Hc} \RB \geq \mathzapf{R} \LB m_i^*, \Cc(r_i, n_i), \underline{\epsilon} \RB -\delta_i$, we choose $r_i^* = \min \Rc(m_i^*)$. The worst case complexity of Algorithm~\ref{alg:la} is $\Oc\LB N_u n_{MCS}\RB$ which is incurred when the for-loop is fully executed, i.e., for each \gls{UE} $i$, $c_i$ is updated sequentially from $k_{max}$ to $1$. This would essentially have tested the \gls{BMDR} criterion for each \gls{MCS} level for each \gls{UE}. For linear detectors, \gls{BMDR} is a non-decreasing function of the post-equalization \gls{SINR}~\cite{kps_jh_part1} which is unaffected for a user by a change of constellation for any other user, provided that all constellations have unit energy. Our numerical studies seem to indicate that \eqref{eq:detector_separable} holds for the $K$-best detector as well. In general, Algorithm~\ref{alg:la} might be suboptimal for arbitrary non-linear detectors, but still is a low-complexity alternative to any other optimal algorithm with higher complexity. 

\begin{algorithm}
  \caption{Pseudocode to select \gls{MCS} for each user based on channel measurements.}  
  \label{alg:la}
  \begin{algorithmic}
    \Input
    \Desc{$\Mc$}{: Set of available QAM modulation orders, $\Mc = \{2k, k=1,\cdots, k_{max} \}$ } 
    \Desc{$\mathcal{R}(m)$}{: Set of available code-rates for  modulation order $m$, $\forall m \in \Mc $} 
    %\Desc{$\underline{\mathscr{R}}$}{: Set of precomputed target \glspl{BMDR} for every available maximum length code  \\ ~~~~~~~~~~~~~~ and a target \gls{CER} $\underline{\epsilon}$. $\underline{\mathscr{R}} = \LP \mathzapf{R}\LB m, \Cc(r, n_{max}), \underline{\epsilon} \RB ~\vert ~ \forall m \in \Mc, r \in \mathcal{R}(m) \RP$}
    \Desc{$\mathzapf{R}$}{: Learned function to predict the target \gls{BMDR} for a target \gls{CER} $\underline{\epsilon}$  given \\ ~~~~~~~~~~~~~~ code-rate $r$, code-length $n$, modulation order $m$}
    \Desc{$\delta_i$}{: Precomputed \gls{BMDR}-correction offset for \gls{UE} $i$, $\forall i = 1,\cdots, N_u$}
    \Desc{$\widehat{\Hc}$}{: Set of estimated composite MU-MIMO channel matrices of size $n_r \times N$}
    \Desc{$n_{RE}$}{: Number of \glspl{RE} for codeword transmission}
    \Desc{$f_{\Dc, i}$}{: A pretrained \gls{BMDR}-predictor for detector $\Dc$ and \gls{UE} $i$, $\forall i = 1,\cdots, N_u$, \\ ~~~~~~~~~~~~~~ takes as inputs the channel matrix and set of modulation orders for all users}  
    \EndInput
    \Output
    \Desc{$mcs_i$}{: Chosen \gls{MCS} index for \gls{UE} $i$, $\forall i = 1,\cdots, N_u$}      
    \EndOutput     
  \Initialize{\strut$\mathrm{updateFlag} \gets \mathrm{True}$ \\ \strut$c_i \gets k_{max} $, $\forall i=1,\cdots,N_u$ \Comment{Begin with the highest modulation order}} 

  \While{$\mathrm{updateFlag}$ is $\mathrm{True}$}
     \State $m_i \gets 2c_i$, $n_i \gets n_{RE} m_in_t^{(i)}$, $\forall i=1,\cdots,N_u$ 
     \Comment{Assign modulation order to user}
     \State $\mathfrak{m} \gets \{m_1,\cdots,m_{N_u} \}$
     \State $\hat{R}_{\Dc,i}\LB \mathfrak{m},\widehat{\Hc}\RB \gets \frac{1}{\left \vert \widehat{\Hc} \right \vert}\sum_{\widehat{\Hm} \in \widehat{\Hc}}f_{\Dc, i}\LB \mathfrak{m},\widehat{\Hm}\RB $,  $\forall i=1,\cdots, N_u$
     \Comment{Estimate average \gls{BMDR}}
     \ForAll{$i=1,\cdots, N_u$}
      \If{$\hat{R}_{\Dc,i}\LB \mathfrak{m},\widehat{\Hc}\RB < \min_{r_i\in \Rc(m_i)} \LP \mathzapf{R} \LB m_i, \Cc(r_i, n_i), \underline{\epsilon} \RB  \RP - \delta_i$}
      \Comment{BMDR criterion}
        \State $c_{i,next} \gets \max(c_i-1,1)$
        \Comment{Reduce modulation order}
      \EndIf
     \EndFor 
     \If {$c_{i,next} = c_i$, $\forall i=1,\cdots, N_u$}
       \State $\mathrm{updateFlag} \gets \mathrm{False}$ \Comment{No more updates of modulation orders}
     \EndIf
     \State $ c_i \gets c_{i,next}$,  $\forall i=1,\cdots, N_u$
  \EndWhile
  \State $m_i^* \gets m_i,~~r_i^* \gets \argmax{r_i \in \mathcal{R}(m_i)} \LP \hat{R}_{\Dc,i}\LB \mathfrak{m},\widehat{\Hc} \RB \geq \mathzapf{R} \LB m_i, \Cc(r_i, n_i), \underline{\epsilon} \RB -\delta_i \RP$, $\forall i=1,\cdots, N_u$
  \State $mcs_i \gets \mathrm{MCS}(m_i^*, r_i^*)$, $\forall i=1,\cdots, N_u$ \Comment{Map $(m_i^*, r_i^*)$ to MCS index}
  \end{algorithmic}
\end{algorithm}

\subsection{Dynamic Detector Selection} \label{subsec:detector_selection}
Suppose that there are $N_d$ available detectors $\LP \Dc_p\RP_{p=1}^{N_d}$ to choose from, with the complexity of $\Dc_p$ denoted by $\mathfrak{C}(\Dc_p)$. We emphasize here that the complexity could either be in terms of time taken to perform detection for one \gls{RE}, or in terms of the number of computations performed during the course of detection for one \gls{RE}. The two need not be proportional to one another, an example being the case where a particular detector performs more operations than its rival but allows parallel processing while its rival does not. Depending on the application, the complexity metric is chosen such that a less "complex" detector is more desirable for usage than its rival if both the detectors are equally reliable. The list of detectors $\{\Dc_1,\cdots,\Dc_{N_d}\}$ is assumed to be ordered so that $\mathfrak{C}(\Dc_1) \leq \cdots \leq \mathfrak{C}(\Dc_{N_d})$. The goal is to choose the least complexity detector that can meet certain target metrics (like throughput or error rates). 

This problem has been considered in various settings in the literature. In \cite{ketonen09}, the authors consider a single user $2\times2$ and $4\times4$ \gls{MIMO} system and compare the performances of three detectors: the \gls{LMMSE} detector, \gls{SIC}, and the $K$-best detector. Through experiments, they recommend the usage of the \gls{LMMSE} detector at low \gls{SINR} and the $K$-best at high \gls{SINR}. In \cite{lai09}, an asymptotic \gls{BER} analysis is performed to obtain \gls{SINR} thresholds that allow detector switching between \gls{SIC} and $K$-best detection. The problem of adaptively allocating computational resources to the detection problem for each channel realization, under a total per codeword complexity constraint, is considered in \cite{Cirkic2011}. Placing the \gls{MLD} on one end of the spectrum and the \gls{ZF} detector at the other end, the paper considers the \gls{MI} of the \gls{MIMO} channel for a Gaussian-distributed input to be the metric for the \gls{MLD} and a similarly calculated \gls{MI} to be the corresponding metric for the \gls{ZF} detector, and proposes a linear interpolation between these two metrics as the metric for any detector with intermediate computational complexity. Since practical communication systems use QAM constellations, the MI for a Gaussian-distributed input is not the ideal choice of metric. Moreover, the aforementioned linear interpolation is not known to be accurate in general. A \gls{NN}-based approach to switch between detectors in each \gls{RE} of an \gls{OFDM}-based \gls{MIMO} system is presented in \cite{Chaudhari2020} while a \gls{RL}-based approach is considered in \cite{Kwon2020}. The limitation of all these works is that the proposed techniques cannot be extended to a \gls{MU-MIMO} setting where different \glspl{UE} can have different \gls{MCS} levels. 

Let $m_{\Dc_p,i}^*$ and $r_{\Dc_p,i}^*$ respectively denote the estimated modulation order and code-rate for \gls{UE} $i$ with detector $\Dc_p$ according to Algorithm~\ref{alg:la}. Then, one can use a hybrid detection strategy that dynamically chooses $\Dc_{p^*}$ where
\begin{align}\label{eq:detector_sel}
p^* & = \min \LP \argmax{p \in \{1,\cdots,N_d\}}\LP \frac{1}{N_u} \sum_{i=1}^{N_u} r_{\Dc_p,i}^*m_{\Dc_p,i}^*   \RP \RP.
\end{align}
In other words, \eqref{eq:detector_sel} attempts to choose the least complexity detector among the ones that offer the best average \gls{SE} subject to a target \gls{CER} constraint for each user. This is a more natural way of selecting a detector in an \gls{MU-MIMO} setting with \gls{LA} compared to the existing works in the literature. Another possible detector selection strategy is to choose $\Dc_{p^*}$ with  
\begin{align}\label{eq:detector_sel1}
  p^* & = \argmax{p \in \{1,\cdots,N_d\}}\LP \gamma f_1(\Dc_p) + (1-\gamma)f_2(\Dc_p)  \RP, 
  \end{align}
where $\gamma \in \LSB 0,1 \RSB$, $f_1(\Dc_p) \triangleq (1/(N_ur_{max}m_{max}))\sum_{i=1}^{N_u} r_{\Dc_p,i}^*m_{\Dc_p,i}^*$, and $f_2(\Dc_p) \triangleq -\mathfrak{C}(\Dc_p)/\mathfrak{C}(\Dc_{N_u})$. Here, $r_{max}m_{max}$ denotes the \gls{SE} corresponding to the highest available \gls{MCS} level. By varying the value of the weight $\gamma$, \eqref{eq:detector_sel1} helps obtain a balance between \gls{SE} and computational complexity. A value of $\gamma$ closer to $1$ puts more emphasis on maximizing the \gls{SE} while a value closer to $0$ puts more emphasis on choosing the least complexity detector. 

%% file: tex/PLA.tex
\section{Physical Layer Abstraction}
\label{sec:pla}

With the system model as described in Section~\ref{sec:system_model}, assume that \gls{UE} $i$ transmits the codeword bits on $n_{RE}$ \glspl{RE}, each \gls{RE} indexed by the frequency-time pair $(f,t)$. Let $\Gc_i$ denote the set of index pairs with the associated set of channel realizations $\Hc_{\Gc_i} = \LP \Hrm_{f,t}, (f,t) \in \Gc_i\RP$. Assuming a linear receiver, the crucial step in \gls{PHY} abstraction in the literature \cite{phy_abs_2005, phy_abs_2006,lagen2021new} is to obtain an effective \gls{SINR} $\bar{\rho}_{i} $ as follows.
\begin{equation}\label{eq:phy_eff_sinr}
    \bar{\rho}_{i} = \beta_1I^{-1} \LB \frac{1}{n_{RE}} \sum_{(f,t) \in \Gc_i} I\LB \frac{\rho_{f,t,i}}{\beta_2}\RB \RB
\end{equation}
where $I()$ is a model-specific function, $\rho_{f,t,i}$ is the post-equalization \gls{SINR} for \gls{UE} $i$ in the \gls{RE} indexed by $(f,t)$, and $\beta_1, \beta_2$ are parameters that allow the model to adapt to the characteristics of the considered \gls{MCS}. \gls{CESM} corresponds to $I(x) = \log{1+x}$, \gls{EESM} to $I(x) = e^{-x}$, \gls{LESM} to $I(x) = \mathrm{log}_{10}(x)$, and \gls{MIESM} to $I(x) = \Ic_m(x)$ where $\Ic_m(x)$ refers to the \gls{MI} between the input and the output in an \gls{AWGN} channel with $2^m$-QAM and an \gls{SNR} of $x$.
Following the computation of this effective \gls{SINR}, a \gls{SINR}-\gls{CER} mapping is performed to obtain the error metrics. Currently, there is no clear consensus on which model-specific function to use, and more importantly, on how to use the above technique for non-linear receivers.

\begin{figure}
    \centering       
        \includegraphics[scale = 0.5]{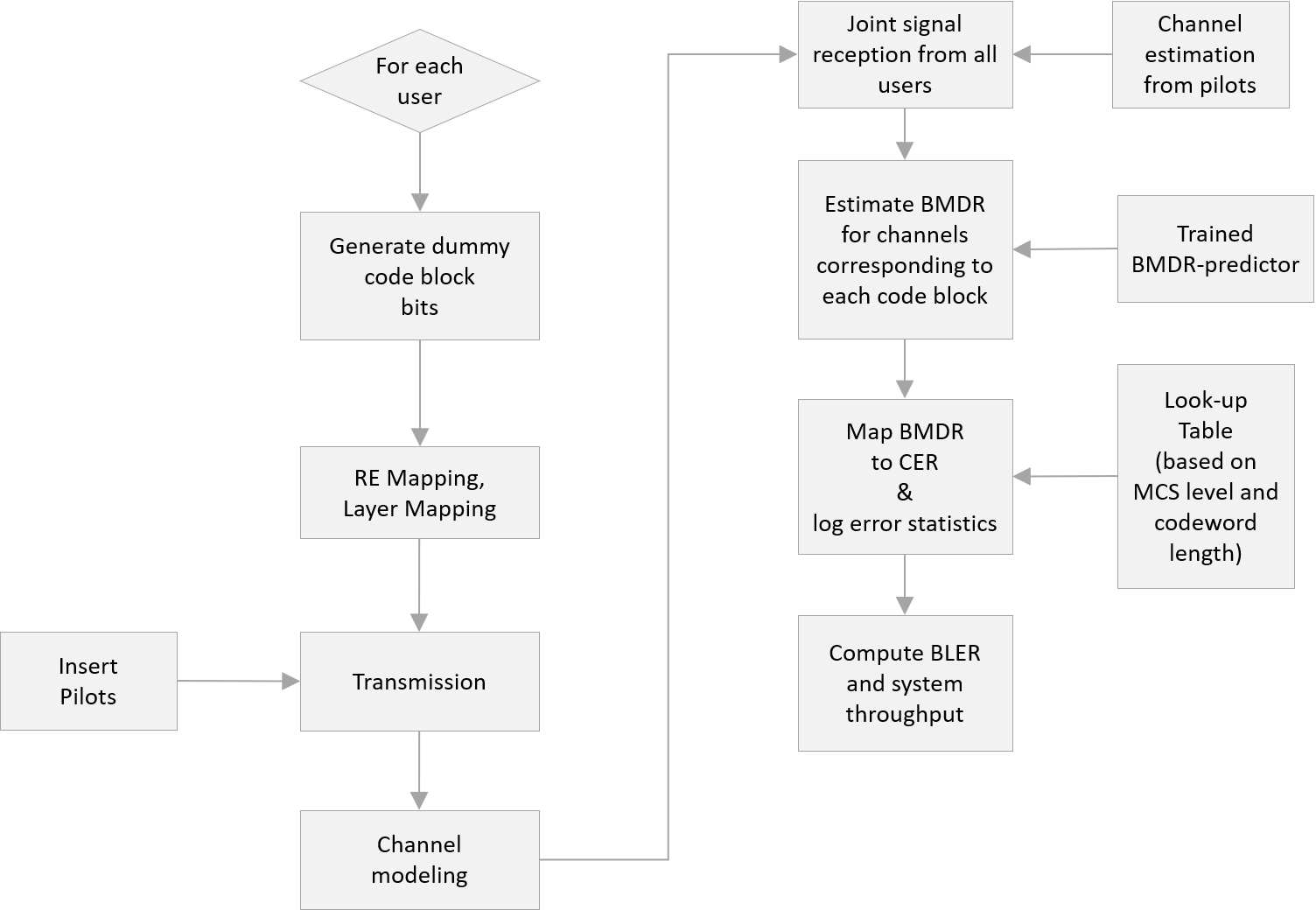}
        \caption{Flowchart of the proposed \gls{PHY} abstraction with the time-intensive procedures of Fig. \ref{fig:pla_flowchart_LLS} replaced by less complex procedures.}
        \label{fig:pla_abstraction}
\end{figure}

\gls{BMDR} serves as a more natural metric for performing \gls{PHY} abstraction. This is because the \gls{BMDR}-\gls{CER} relationship is clearer than the \gls{ESM}-\gls{CER} relationship, where the effective \gls{SINR} is calculated using \eqref{eq:phy_eff_sinr}. We note that while \gls{BMDR} is equivalent to \gls{MIESM} for linear receivers, it can be applied to non-linear receivers as well. Our proposed approach is to calculate 
\begin{equation}\label{eq:phy_eff_bmdr}
    \hat{R}_{\Dc,i}\LB \mathfrak{m}, \Hc_{\Gc_i} \RB = \frac{1}{n_{RE}}\sum_{(f,t) \in \Gc_i} \hat{R}_{\Dc,i}\LB \mathfrak{m}, \Hrm_{f,t} \RB
\end{equation} 
where $\mathfrak{m} = \LP m_1,\cdots,m_{N_u} \RP$ is the set of modulation orders used by all the co-scheduled users, and then map $\hat{R}_{\Dc,i}\LB \mathfrak{m}, \Hc_{\Gc_i} \RB$ to a \gls{CER} using the lookup table $\Lc_U$ whose construction was detailed in Section \ref{sec:bmdr}. With respect to \gls{5GNR} terminology~\cite{3GPP_coding_2020}, suppose that \gls{UE} $i$ notionally transmits $B$ \glspl{TB}, with each \gls{TB} being segmented into $L$ \glspl{CB} (a \gls{CB} is a block of message bits). Therefore, a total of $BL$ codewords are transmitted notionally (not actually transmitted because of the abstraction model).  Then, the \gls{BMDR} associated with each codeword transmission is estimated and mapped to a \gls{CER} value from the lookup table. Let the estimated \gls{BMDR} associated with the $l^{th}$ codeword of \gls{TB} $j$ be $\hat{R}^{(j,l)}_{\Dc,i}\LB \mathfrak{m}, \Hc_{\Gc_i}^{(j,l)} \RB$ where $\Hc_{\Gc_i}^{(j,l)}$ is the associated set of channel realizations, and let  $\hat{p}_{i,j,l}$ denote the estimated CER for this codeword.  Then, 
\begin{equation}
    \hat{p}_{i,j,l} = \argmin {P_e(R_{MLD}) \in \Lc_U}\LP \left \vert R_{MLD} - \hat{R}^{(j,l)}_{\Dc,i}\LB \mathfrak{m}, \Hc_{\Gc_i}^{(j,l)} \RB \right \vert  \RP
\end{equation}
and the estimated probability of error for \gls{TB} $j$ is $\hat{P}_{i,j} = 1 - \prod_{l=1}^{L}\LB 1 - \hat{p}_{i,j,l}\RB$. The overall \gls{BLER} for \gls{UE} $i$ is $\frac{1}{B}\sum_{j=1}^B\hat{P}_{i,j}$. The average system throughput can be estimated from the \glspl{BLER} of all the users. Fig.~\ref{fig:pla_abstraction} shows the flowchart of the proposed technique. 

%% file: tex/simulation_results.tex
\section{Simulation Results}
\label{sec:sim_results}

\subsection{Simulation Setup} \label{subsec:sim_setup}
\begin{table}[htbp]
    \begin{center}
        \begin{tabular}{|c|c|}
        \hline
            {\bf Parameter}      & {\bf Value}                         \\ \hline \hline
            $N_u$  & $4$    \\ \hline
            $n_t^{(i)}$   & $1$, $\forall i = 1,\cdots, 4$  \\ \hline
			$n_r$  & $16$ ($2$H$\times 8$V) \\ \hline
            Number of cells in grid  & $21$ \\ \hline
			Total number of users in grid & $210$ \\ \hline
            Inter-site distance   & \SI{200}{\metre}   \\ \hline
            Channel model  & 38.901 Urban Micro (UMi) NLoS  \\ \hline
            Carrier frequency  & \SI{3.5}{\GHz}   \\ \hline              
            \gls{OFDM} subcarrier spacing   & \SI{30}{\kHz} \\ \hline
			System bandwidth  & \SI{8.64}{\MHz}   \\ \hline          
            Number of OFDM symbols/slot& $14$ \\ \hline
			\gls{OFDM} symbol duration &\SI{35.7}{\microsecond} \\ \hline
            Number of \glspl{PRB}/slot& $24$ \\ \hline
            \gls{UE} speed   &\SI{5}{\kmph}  \\ \hline
            \gls{UE} max transmit power $P_{max}$  & \SI{23}{\dBm} \\ \hline
			\gls{UE} uplink power control $(P_0, \alpha)$  & (\SI{-98}{\dBm}, $0.7$), (\SI{-98}{\dBm}, $1$)   \\ \hline
        \end{tabular}
    \end{center}
    \caption{A list of simulation parameters.}
    \label{tab:parameters}
\end{table}

\begin{figure}
	\centering			
		\includegraphics[scale = 0.8]{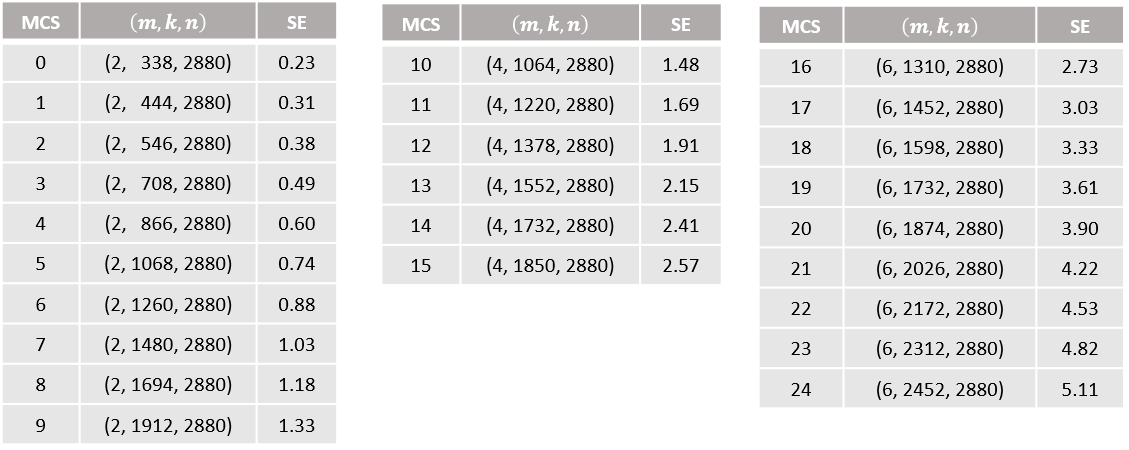}	
		\caption{The \gls{MCS} table used for simulations.}
		\label{fig:mcs_table}
\end{figure}

We consider the following setup (with a summary of the parameters in Table~\ref{tab:parameters}) for performing multi-cell, multi-link-level simulations, the code for which was written in NumPy (for channel generation) and TensorFlow. We consider 7 sites with 3 cells per site, leading to a total of $21$ cells arranged in a hexagonal grid with wraparound (which essentially means that each cell sees an inter-cell interference pattern similar to that of the central cell). The inter-site distance is \SI{200}{\metre}, and we consider the 38.901 Urban Micro (UMi) NLoS~\cite[Section 7.2]{3GPP_ch_model_2019} channel model. A total of $210$ users are dropped at random in this grid. The carrier frequency is \SI{3.5}{\GHz} and each \gls{BS} is equipped with a rectangular planar array consisting of $16$ ($2$ vertical, $8$ horizontal) single-polarized antennas installed at a height of \SI{25}{\metre} and an antenna spacing of $0.5 \lambda$, where $\lambda$ is the carrier wavelength. The \glspl{UE} are each equipped with $2$ dual-polarized antennas ($1$ vertical, $2$ horizontal), and the maximum total output power $ P_{max}$ per \gls{UE} is \SI{23}{\dBm}. We consider a regular 5G \gls{OFDM} grid with a subcarrier spacing of \SI{30}{\kHz}, a slot of $14$ symbols (of duration \SI{35.7}{\microsecond} each), and $24$ \glspl{PRB} (of $12$ subcarriers each) leading to a total usable system bandwidth of \SI{8.64}{\MHz}. Out of the $14$ symbols, four are used for \gls{DMRS} transmission and the remaining for data. The total uplink transmit power $\rho$ (in dBm) used by each \gls{UE} is given by the following \gls{OLPC} equation~\cite[Section 7]{3GPP_UL_control_2021}:
\begin{equation}
	\rho = \min\{ P_{max}, P_0 + 10\logt{N_{PRB}} + \alpha PL \}
\end{equation}
where $ P_{max} = 23$, $N_{PRB} = 24$ is the total number of \glspl{PRB}, $PL$ is the pathloss estimate (based on the measured channel gains on the downlink), $P_0$ is the expected received power per \gls{PRB} under full pathloss compensation, and $\alpha \in [0,1]$ is the fractional pathloss compensation factor. In our simulations, we take $P_0$ to be \SI{-98}{\dBm} and $\alpha \in \{0.7, 1\}$. The transmission of each codeword is completed within each slot, so we fix the codeword length to be $2880$ for all \gls{MCS} levels. A custom \gls{MCS} table as shown in Fig.~\ref{fig:mcs_table} is used, where $m$ refers to the modulation order, $k$ to the message length, $n$ to the codeword length, and $\textrm{SE} = mk/n$. 5G \gls{LDPC} codes are used for channel coding.

Once the users are dropped, they are assumed to have a speed of \SI{5}{\kmph}, and the simulations are evaluated for $500$ slots (of \SI{0.5}{\ms} each). The metrics are evaluated for ten such independent user drops. The number of transmitting layers is fixed to be four; so, depending on the users' positions, a single user can transmit using all of its four antennas, or four users can transmit using a single antenna each (and any combination of user antennas summing to four). The users are scheduled as follows: for each user, its serving cell is identified to be the one for which the average channel gains are the highest. Next, the $n_r \times n_r$ sized channel covariance matrix for each user is estimated at the \gls{BS} using the \gls{SRS} signals sent by the user, and is calculated by averaging over both time and frequency. Then, the set of users in each cell is divided into (not disjoint) subsets such that in each subset, the cross-correlation between the principal eigenvectors of the channel covariance matrices of any two users is less than $0.1$. The cardinality of each subset is limited to four, and the subsets are served in a round-robin fashion. \gls{LMMSE} channel estimation is performed for each user, and the interference covariance matrix at each cell is assumed to be perfectly known (i.e., with the notation as used in Section~\ref{sec:system_model}, $\Km_n$ is assumed to be perfectly known while $\Km_e$ is estimated). 

The \gls{BMDR}-\gls{CER} map is generated as explained in Section~\ref{sec:bmdr} using $100,000$ codewords for each \gls{SNR}, and the \gls{BMDR} is computed using $10,000$ independent input symbol and noise realizations for the same \gls{SNR} in order to do the Monte-Carlo approximation of \eqref{eq:gen_bmdr_def} for the \gls{SISO}-\gls{AWGN} channel.

For \gls{LLR} generation, we consider the \gls{LMMSE} detector as our choice of linear detector, and the $32$-best detector ($K=32$ in~\cite{Guo2006}) as our choice of non-linear detector. Their respective \gls{BMDR}-predictors are trained as explained in \cite[Section V]{kps_jh_part1}. We would like to reiterate that the reason for choosing the $32$-best detector is due to its relatively low complexity compared to other non-linear techniques. In practice, it is often required to go beyond $K=64$ in order to achieve significant performance gains over \gls{LMMSE}. The purpose of our simulations is not to show that one detector is superior to the other, but rather to corroborate our claims about the role of \gls{BMDR} in \gls{LA} and \gls{PHY} abstraction for both linear and non-linear detectors. 

The throughput for each user in a given user drop, when using a particular detector, is obtained as follows: Suppose that \gls{UE} $i$ transmits a total of $L_i$ codewords in that drop, with the $j^{th}$ codeword carrying $k_{ij}$ message bits. Let the total number of slots for that drop be $T$, with each slot being of duration $t_{slot}$ seconds. Let $\delta_{ij} = 1$ if the $j^{th}$ codeword is correctly decoded using the \gls{LDPC} decoder for which the input \glspl{LLR} are generated by the detector in context, and $0$ otherwise. {\it We do not consider any type of \gls{ARQ} in our simulations}. Then, the throughput $TP_i$ in megabits per second (Mbps) for \gls{UE} $i$ in that user drop is
\begin{equation}
	TP_i = \frac{1}{10^{6}}\LB\frac{\sum_{j=1}^{L_i}\delta_{ij}k_{ij}}{Tt_{slot}}\RB. 
\end{equation}
The \gls{AM} and the \gls{GM} of the throughputs for all the users in a drop are then recorded, and this process is repeated for each of the ten independent drops. While the throughput for each user is heavily dependent on the user-scheduling algorithm, the goal here is to assess the performance of the detectors considered. 

\subsection{\gls{LA}: Numerical Results} \label{subsec:sim_LA}

For \gls{MCS} selection, we use the previous slot's channel realizations even if the users didn't transmit, except for the first slot in which case we estimate the first slot's channels before selecting the \gls{MCS}. We choose a target \gls{CER} of $10^{-3}$. With this setup, we simulate the following four receiver algorithms:
\begin{enumerate}
	\item {\it \gls{LMMSE} detection with \gls{EESM}-based \gls{MCS} selection}: In this case, we use the \gls{LMMSE} detector and use \gls{EESM} for \gls{MCS} estimation. To be precise, we compute the post-equalization \gls{SINR} for each user in each \gls{PRB} using the previous slot's channel realizations. Let $\hat{\rho}_{i,n}$ denote the post-equalization \gls{SINR} for \gls{UE} $i$ in the $n^{th}$ \gls{PRB}, $n=1,\cdots,24$. The   effective \gls{SINR} $\bar{\rho}_{i}$ is estimated to be \cite{Lagen2020} 
	\begin{equation}\label{eq:eff_sinr}
		\bar{\rho}_{i} = -\beta \ln {\frac{1}{24} \sum_{n = 1}^{24} \exp{-\frac{\hat{\rho}_{i,n}}{\beta}} }
	\end{equation}
    where the value of $\beta$ is chosen according to \cite[Table II]{Lagen2020} for each available \gls{MCS}. Finally, the highest \gls{MCS} that meets the target \gls{CER} is selected using an \gls{SNR}-\gls{CER} map. We call this scheme ``LMMSE-EESM".
	\item {\it \gls{LMMSE} detection with \gls{BMDR}-based \gls{MCS} selection}: This uses the \gls{LMMSE} detector with the \gls{MCS} selected according to Algorithm \ref{alg:la}. This scheme is called  ``LMMSE-BMDR".
	\item {\it $32$-best detection with \gls{BMDR}-based \gls{MCS} selection}: This scheme uses the $32$-best detector and Algorithm \ref{alg:la}, and is called ``$32$-best-BMDR''.
	\item {\it Hybrid detector}: selects either the \gls{LMMSE} detector or the $32$-best detector in each slot according to~\eqref{eq:detector_sel}, and is called ``Hybrid-BMDR". 
\end{enumerate}
We have considered the \gls{EESM}-based effective \gls{SINR} mapping because it is the one recommended by 3GPP. We did not explicitly simulate the \gls{RBIR}-based effective \gls{SINR} mapping scheme (which is a variant of the \gls{MIESM}-based mapping and is proposed for usage in IEEE 802.11) because, for linear detectors, this turns out to be equivalent to our proposed \gls{BMDR}-based approach. Since there is no known effective \gls{SINR} mapping scheme for non-linear detectors, we have only considered the proposed \gls{BMDR}-based approach for the $32$-best detector.

\begin{figure}
	\centering			
		\includegraphics[scale = 0.45]{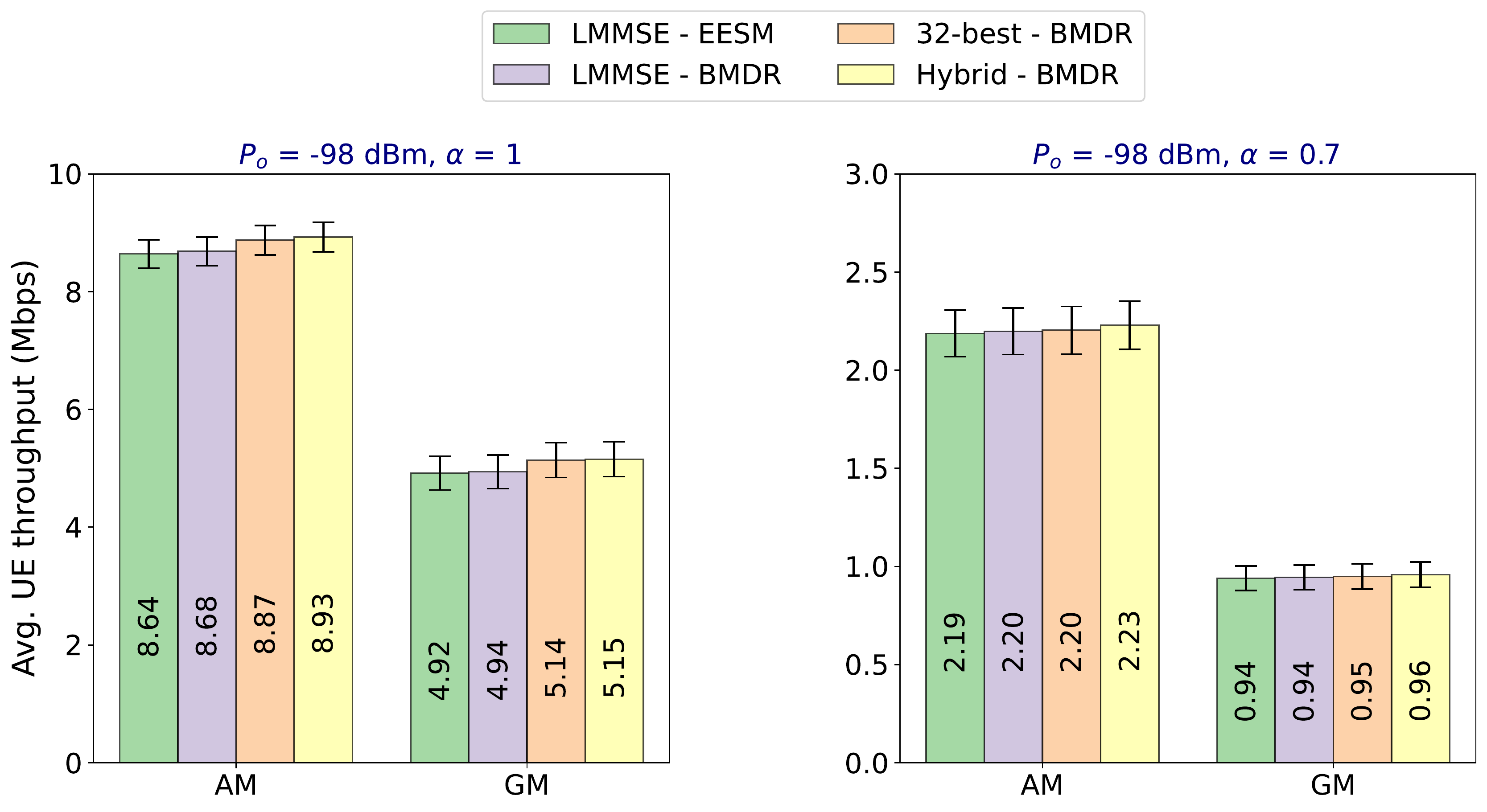}	
		\caption{The \gls{AM} and \gls{GM} throughputs with $90\%$ confidence intervals for the four detection schemes for (left) $\alpha=1$ and (right) $\alpha=0.7$.}
		\label{fig:mean_tput}
\end{figure}

\begin{figure}
	\centering			
		\includegraphics[scale = 0.45]{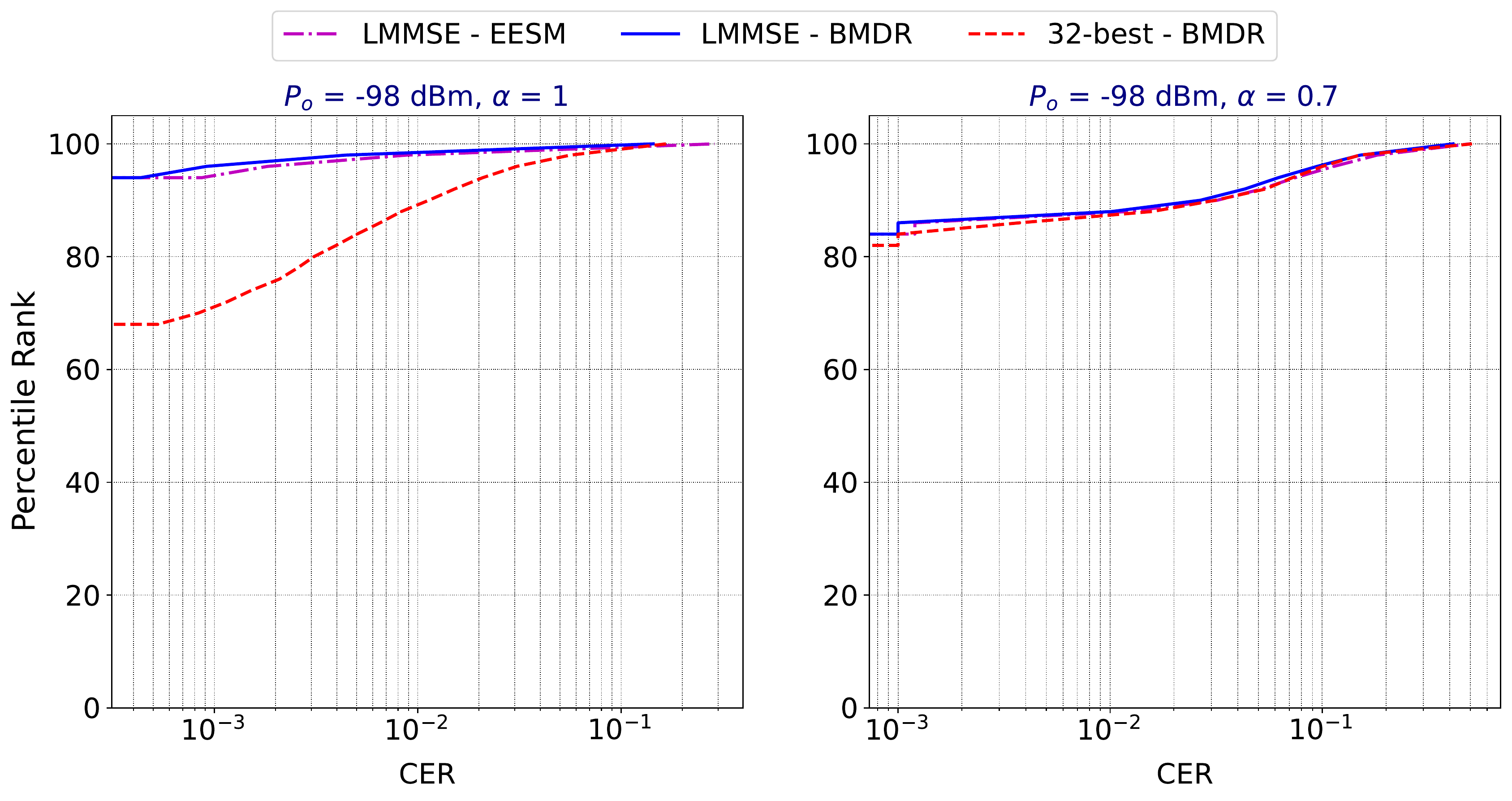}
		\caption{The percentiles of the \gls{CER} for the \gls{LMMSE} detector and the $32$-best detector for (left) $\alpha=1$ and (right) $\alpha=0.7$. For any \gls{CER} value $p_e$ on the X-axis, the corresponding value on the Y-axis indicates the percentage of users (across all drops) that achieved a \gls{CER} $ \leq p_e$.}
		\label{fig:percentile_cer}
\end{figure}

\begin{figure}
	\centering			
		\includegraphics[scale = 0.45]{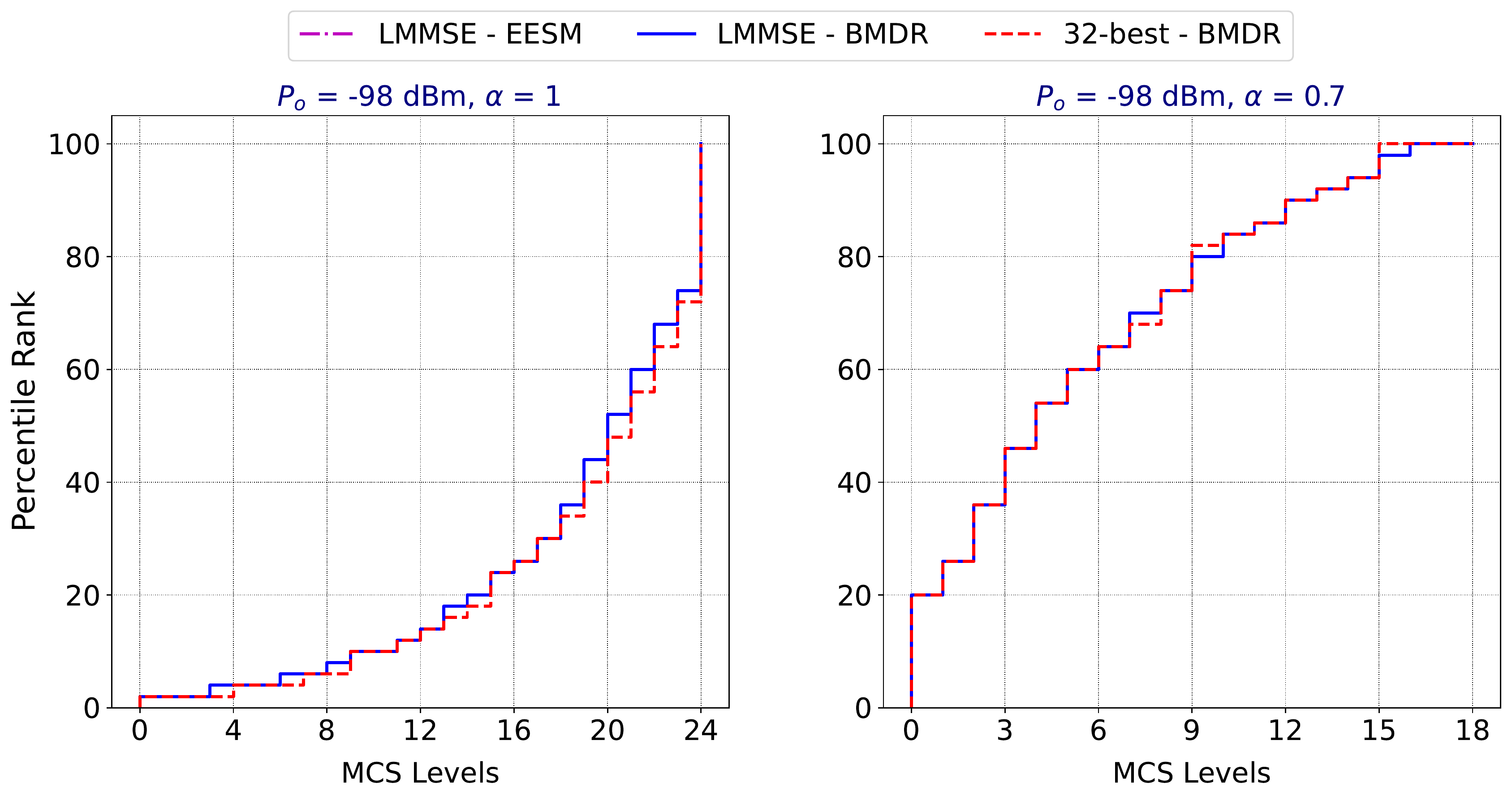}
		\caption{The percentiles of the \gls{MCS} levels for the \gls{LMMSE} detector and the $32$-best detector for (left) $\alpha=1$ and (right) $\alpha=0.7$.}
		\label{fig:percentile_mcs}
\end{figure}

\begin{figure}
	\centering			
		\includegraphics[scale = 0.45]{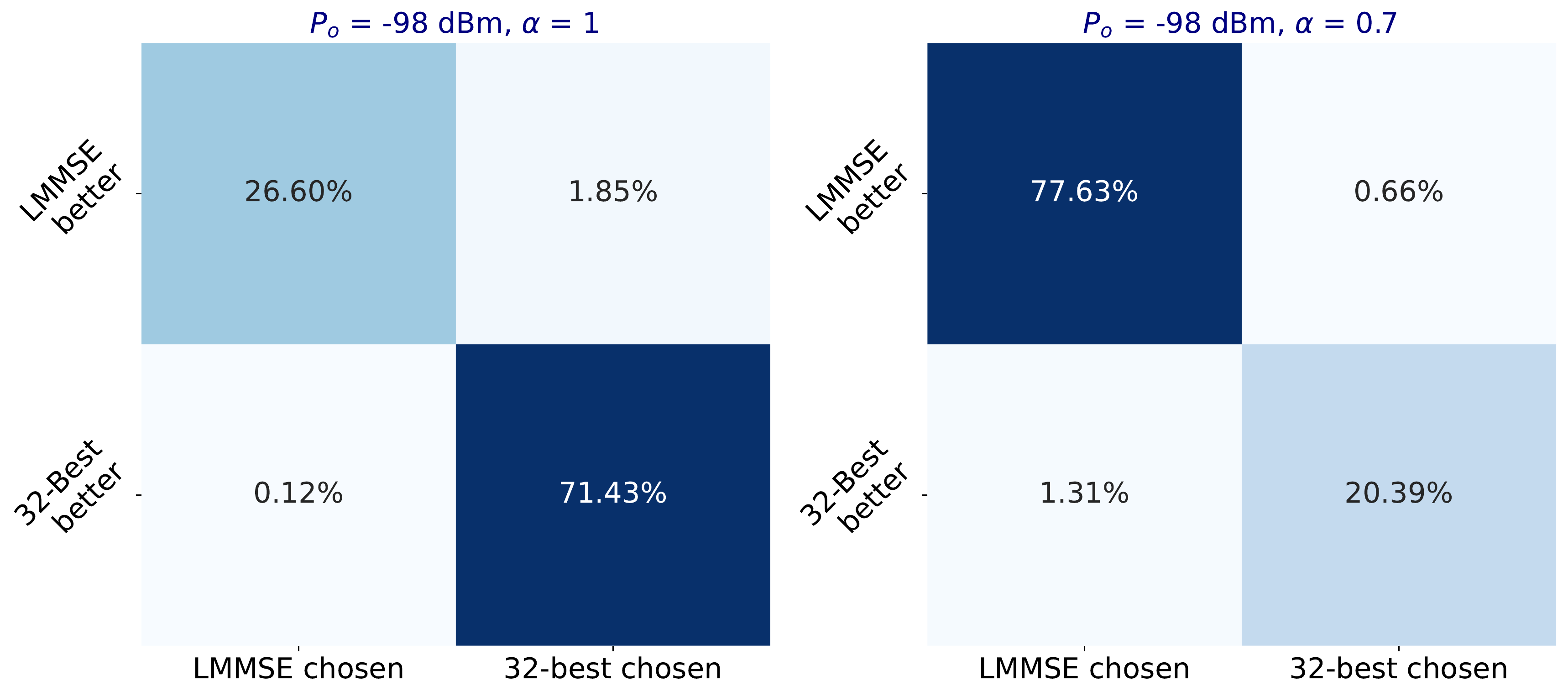}	
		\caption{The confusion matrix for detector selection using the proposed technique for (left) $\alpha=1$ and (right) $\alpha=0.7$.}
		\label{fig:CM}
\end{figure}

Fig.~\ref{fig:mean_tput} shows the plots of the average \gls{AM} and \gls{GM} throughputs for the four schemes along with the $90\%$ confidence intervals (for different user drops). Fig.~\ref{fig:percentile_cer} shows the percentiles of the \gls{CER} for the \gls{LMMSE} detector (both \gls{EESM}-based and \gls{BMDR}-based) and the $32$-best detector, so that for each \gls{CER} value $p_e$ on the X-axis, the corresponding value on the Y-axis indicates the percentage of users (across all drops) that achieved a \gls{CER} less than or equal to $p_e$. Fig.~\ref{fig:percentile_mcs} shows the percentiles of the \gls{MCS} levels selected by the different schemes.

It can be seen from Fig.~\ref{fig:mean_tput} that the hybrid detector gives the best performance, while LMMSE-EESM and LMMSE-BMDR have nearly the same performance. This is not surprising, given that both \gls{EESM} and \gls{MIESM} have been thoroughly evaluated in the literature to provide near-optimal performance, and the proposed \gls{BMDR}-based approach for \gls{LMMSE} detection is equivalent to the \gls{MIESM}-based approach. From Fig.~\ref{fig:percentile_cer}, it can be seen that for the case of $\alpha =1$ (full path-loss compensation), users achieve the target \gls{CER} of $10^{-3}$ with LMMSE detection (both \gls{EESM}-based and \gls{BMDR}-based) around $94 \%$ of the time while this number is $68\%$ for the $32$-best detector. The reason for this difference, as highlighted in \cite{kps_jh_part1}, is the ability to perform more accurate \gls{BMDR}-prediction for a linear detector than for a non-linear detector, and is down to the accuracy of the trained \gls{BMDR}-predictors. For the case of $\alpha = 0.7$ (with partial path-loss compensation), the percentage of the time the \gls{CER} target is met is around $82\%$ for both the detectors. This is due to an increase in the usage of the lowest available \gls{MCS}, as observed in Fig.~\ref{fig:percentile_mcs}. Since the path-loss is not fully compensated, it might be that some users might not have sufficient \gls{SINR} for error-free transmission even at the lowest \gls{MCS} level. From Fig.~\ref{fig:percentile_mcs}, it can be readily seen that the users experience a wide range of \gls{MCS} values for all detection schemes, and depending on the path-loss compensation factor, have a bias towards either the lower \gls{MCS} levels ($\alpha=0.7$) or the higher ones ($\alpha = 1$). For the case of $\alpha=1$, the plots for the \gls{LMMSE} detector with \gls{EESM}-based and \gls{BMDR}-based mapping coincide, and the $32$-best detector allows slightly higher \gls{MCS} selection, evidenced by its curve being below that of the \gls{LMMSE} detector.

 Fig.~\ref{fig:CM} shows the confusion matrices of the detector selection algorithm which corroborates the efficacy of the proposed method. Since the $32$-best detector uses only $32$ survivors per layer in the search tree, there are scenarios where the \glspl{LLR} might be overestimated due to the lack of a sufficient number of candidates of the opposite polarity for each transmitted bit. This leads to poorer performance compared to LMMSE in such cases. {\it Therefore, depending on the channel conditions, the hybrid detection scheme allows us to use a combination of a low-complexity detector and higher complexity non-linear detector for an overall better performance}. To summarize, the main inferences from Figs.~\ref{fig:mean_tput}--\ref{fig:CM} are:
\begin{enumerate}
	\item The proposed \gls{MCS} selection algorithm based on \gls{BMDR} works as well as the state-of-the-art technique that is based on \gls{EESM} for the \gls{LMMSE} detector (Figs.~\ref{fig:mean_tput}--\ref{fig:percentile_cer}).
	\item The proposed \gls{MCS} selection scheme for the $32$-best detector based solely on \gls{BMDR} is quite effective in that most of the users meet the target \gls{CER} (Fig.~\ref{fig:percentile_cer}), and allows slightly better \gls{MCS} selection than the \gls{LMMSE} detector (the left plot of Fig.~\ref{fig:percentile_mcs}).  
	\item The proposed hybrid detector selection algorithm is reasonably accurate in choosing the better detector (Fig. \ref{fig:CM}).
\end{enumerate}

\subsection{\gls{PHY} Abstraction: Numerical Results}

\begin{figure}
	\centering			
		\includegraphics[scale = 0.38]{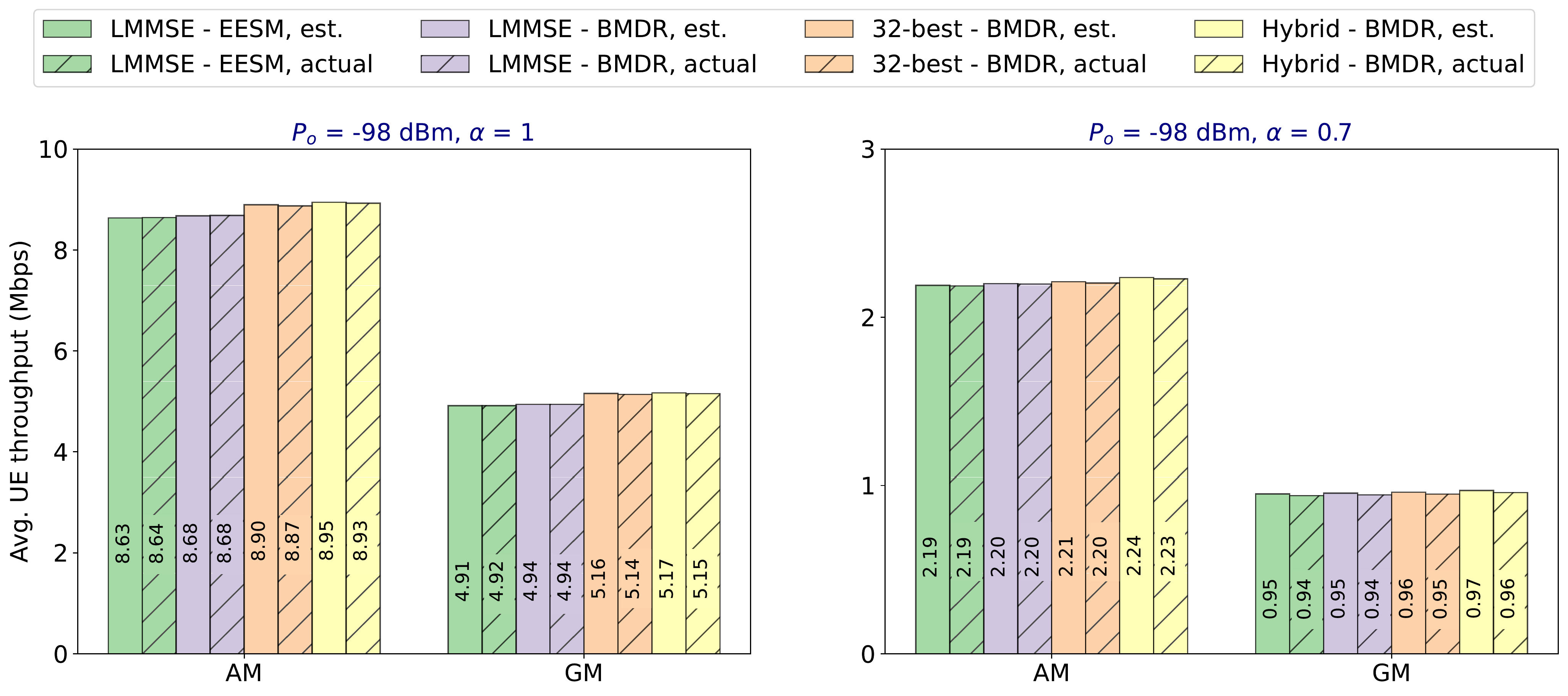}
		\caption{The \gls{AM} and \gls{GM} for the four detector schemes using \gls{PHY} abstraction (compared with the actual values corresponding to Fig.~\ref{fig:mean_tput}) for (left) $\alpha=1$ and (right) $\alpha=0.7$.}
		\label{fig:PLA_mean_tput}
\end{figure}

\begin{figure}
	\centering			
		\includegraphics[scale = 0.45]{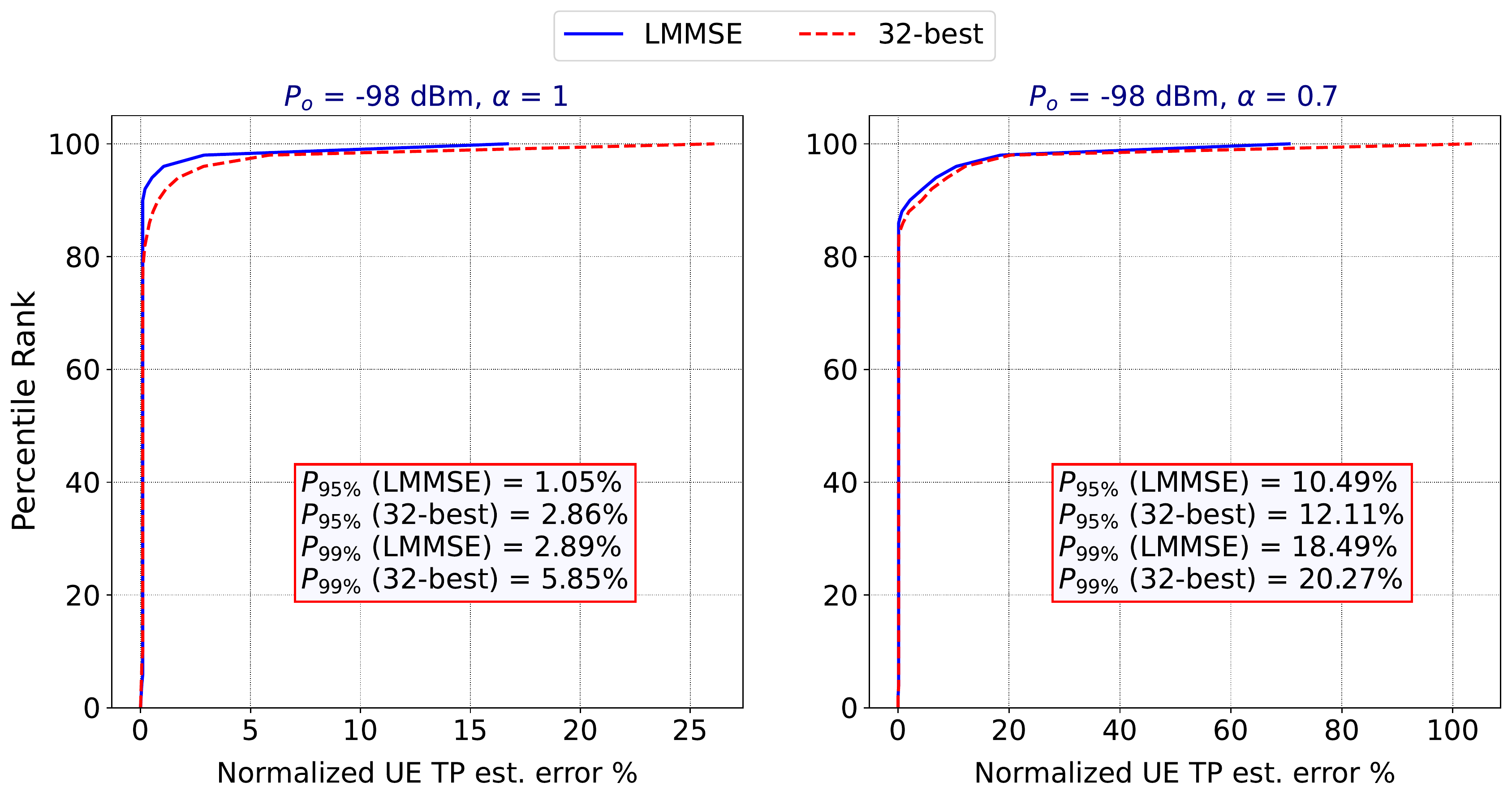}	
		\caption{The percentiles of the normalized \gls{UE}-throughput estimation error using \gls{PHY} abstraction for the \gls{LMMSE} detector and the $32$-best detector for (left) $\alpha=1$ and (right) $\alpha=0.7$.}
		\label{fig:percentile_est_error_PLA}
\end{figure}

\begin{figure}
	\centering	
		\includegraphics[height=0.7\textheight, width=\textwidth]{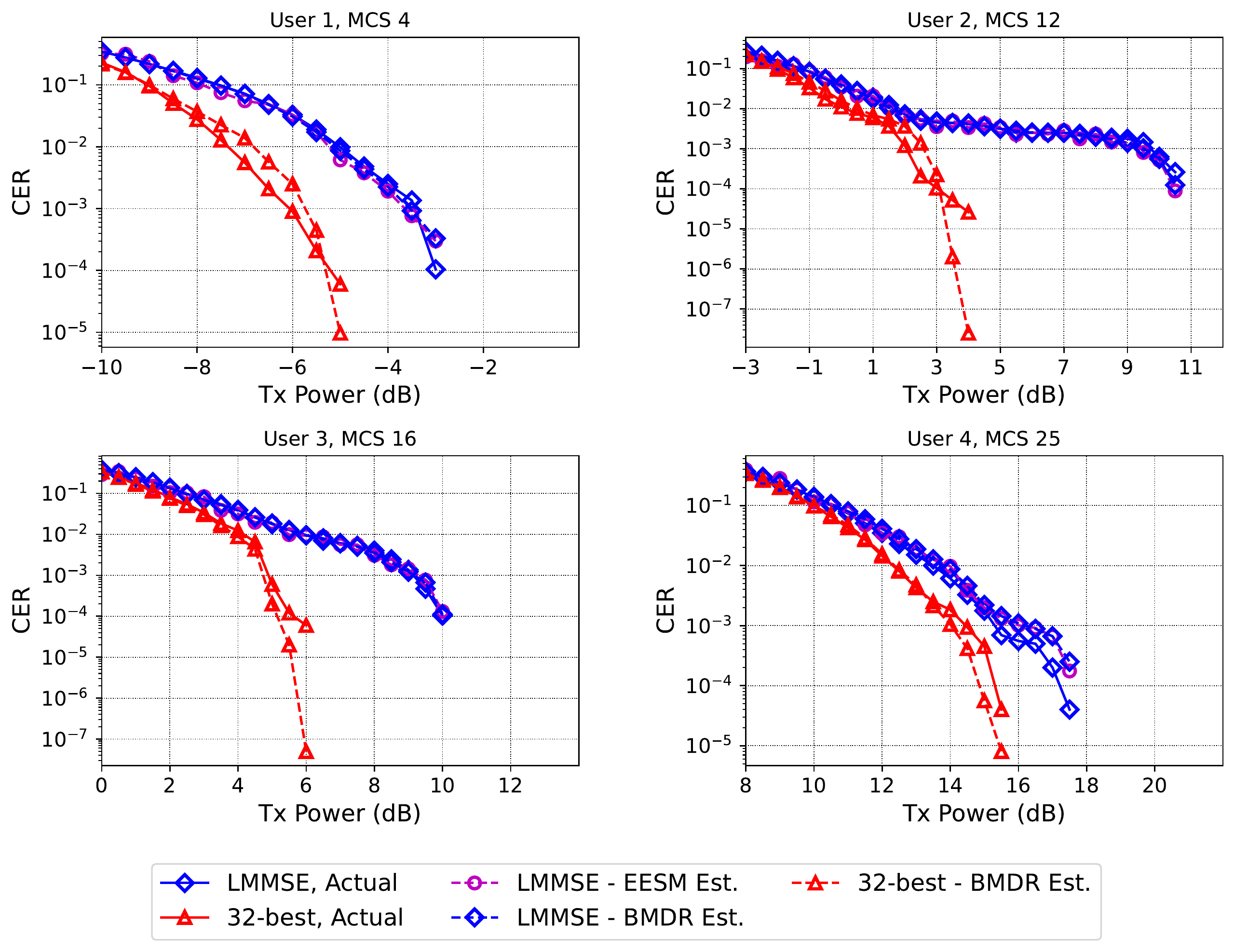}
		\caption{The link-level \gls{CER} v. transmit power (\si{\dB}) plots for the \gls{LMMSE} detector (both \gls{EESM} and \gls{BMDR} based) and the $32$-best detector for four \glspl{UE} with different \gls{MCS} levels. The solid lines correspond to the actual \glspl{CER} and the dashed lines to the \glspl{CER} estimated using the proposed abstraction technique. }
		\label{fig:PLA_LL_CERs}
\end{figure}

We consider the same simulation setup as described in the previous subsection, but do not perform data transmission, \gls{LLR}-generation, or channel decoding. We follow the method outlined in the flowchart of Fig.~\ref{fig:pla_abstraction} and compute the \gls{AM} and \gls{GM} throughputs using the \gls{BMDR}-\gls{CER} mapping. To be precise, the \gls{MCS} levels are selected as in the previous subsection, and the estimated \gls{SE} is multiplied by the mapped \gls{CER} for each notional codeword transmission. This is then averaged for all users in all the drops. The computed \gls{AM} and \gls{GM} throughputs are plotted in Fig.~\ref{fig:PLA_mean_tput} alongside the actual values for all the four schemes. The efficacy of the \gls{EESM} metric is well-known in the literature for linear detectors. Fig.~\ref{fig:PLA_mean_tput} indicates that the \gls{BMDR}-based \gls{PHY} abstraction is equally effective for both linear and non-linear detectors. 

The percentiles of the normalized \gls{UE}-throughput estimation error are shown in Fig.~\ref{fig:percentile_est_error_PLA} for the LMMSE and the $32$-best detectors with \gls{BMDR}-based estimation. We haven't included \gls{LMMSE}-\gls{EESM} in these plots because the corresponding plots are indistinguishable from that of \gls{LMMSE}-\gls{BMDR}. This normalized \gls{UE}-throughput estimation error is calculated as follows: if the actual \gls{UE} throughput is $TP_{actual}$ and the estimated throughput is $TP_{est}$, the normalized throughput estimation error is $\vert TP_{actual} -  TP_{est} \vert/TP_{actual}$. Also marked in Fig.~\ref{fig:percentile_est_error_PLA} are the $95^{th}$ and $99^{th}$ percentiles of the normalized estimation errors for both \gls{LMMSE} and $32$-best detector. For example, in the case of $\alpha = 0.7$, $P_{99\%}$ for the \gls{LMMSE} detector with \gls{BMDR}-based throughput estimation is $18.49\%$, which means that $95\%$ of the normalized \gls{UE}-throughput estimation errors across all the drops are within $18.49\%$ of the true values, i.e., $\vert TP_{actual} -  TP_{est} \vert/TP_{actual} \leq 0.1849$ for $95\%$ of the \glspl{UE}. The results are much better for $\alpha=1$. The high deviations for the case of $\alpha = 0.7$ are for \glspl{UE} with very low throughput (possibly cell-edge \glspl{UE}). In both the plots, the estimation errors are very low for a majority of the \glspl{UE} for both the detectors. 

In Fig.~\ref{fig:PLA_LL_CERs}, we present the \gls{CER} plots as a function of \gls{UE} transmit power in a link-level simulation setting, with the channels generated using the QuaDRiGa channel-simulator~\cite{Jaeckel2014}. In this plot, we have plotted the actual observed \gls{CER} values (solid lines) along with the estimated \gls{CER} (dashed lines) for the case of both \gls{EESM} and \gls{BMDR}. We consider four users with different \gls{MCS} levels which correspond to the $64$-QAM \gls{MCS} table in \gls{5GNR}~\cite[Table 5.1.3.1--1]{3GPP_MCS_table_2020}. The number of transmitted codewords is limited to $10,000$ and hence, the estimated \glspl{CER} deviate a little at values below $10^{-5}$. The curves for \gls{LMMSE}-\gls{EESM} and \gls{LMMSE}-\gls{BMDR} are nearly indistinguishable. The plots show that the proposed link-level modeling technique is quite useful in capturing the main behavior of the \gls{PHY} components for both linear and non-linear detectors. 

While it might be surprising to note that the $32$-best detector performs significantly better than the \gls{LMMSE} detector in Fig.~\ref{fig:PLA_LL_CERs} while the same is not reflected in Fig.~\ref{fig:mean_tput}, the explanation is the following. The link-level plots were obtained using the QuaDRiGa channel-simulator for a single user drop, and the channels of the four users were highly correlated which implies that the channel matrices had large condition numbers. The condition number of a matrix is defined as the ratio of the largest singular value to the smallest singular value. When the condition number of a matrix, expressed in logarithmic form, exceeds \SI{10}{\dB}, it is well-known that linear detectors perform significantly worse than the sphere-decoding variants~\cite{ketonen09}. The condition numbers of the channels generated by the QuaDRiGa channel-simulator exceeded \SI{10}{\dB} approximately $40\%$ of the time. On the other hand, the plots in Fig.~\ref{fig:mean_tput} were obtained by co-scheduling users with low-channel correlation (as explained in Section \ref{subsec:sim_setup}) for which the \gls{LMMSE} detector performs reasonably well. In the latter scenario, the $32$-best detector can only provide a marginal improvement over the \gls{LMMSE} detector. The other important point to note is that since the number of surviving candidates is restricted to only $32$, the quality of the \glspl{LLR} generated by the $32$-best detector is worse than that of the \gls{LMMSE} detector at low transmit power levels. This effect is more visible in the case of $\alpha = 0.7$ (Figs. \ref{fig:percentile_mcs}--\ref{fig:CM}). In order to see significantly enhanced throughputs for such a case, one might need to use a higher value of $K$ for the $K$-best detector ($K \geq 64$).

\begin{remark}
The whole purpose of link-level modeling is to save time while performing \gls{SLS}. \gls{PHY} abstraction for linear detectors using \gls{BMDR} has a time-complexity similar to that of existing techniques based on \gls{EESM}, \gls{MIESM}, \gls{CESM}, or \gls{LESM}. The proposed technique computes the post-equalization \gls{SINR} and maps it to \gls{BMDR} which is then mapped to a \gls{CER}/\gls{BLER} value. In the case of existing techniques, the post-equalization \gls{SINR} is mapped to an effective \gls{SINR} using the relevant \gls{ESM} metric and a look-up table, and this effective \gls{SINR} is then mapped to a \gls{CER}/\gls{BLER} value. 

For the case of non-linear detectors, there is no existing technique for comparison. The natural question to ask is whether the complexity of the proposed \gls{BMDR} prediction method in \cite{kps_jh_part1} is significantly lower than that incurred by actually performing the entire link-level simulation without abstraction. For non-linear receivers like the sphere-decoding variants, the complexity of generating \glspl{LLR} for the system considered in Section \ref{sec:system_model} is exponential in $N$, where $N$ is the total number of transmitting streams from all the \glspl{UE}. For any other iterative non-linear receiver like \gls{SIC}, the very sequential nature of the algorithm induces time latency. The proposed \gls{BMDR} prediction is machine-learning-based and needs the \glspl{CNN} to be trained offline once. Each \gls{CNN} for a \gls{MU-MIMO} configuration has around $5000$ trainable parameters and any real-time usage entails low time-latency because there are no iterations. 
\end{remark}

%% file: tex/discussion.tex
\section{Discussion and Concluding Remarks}
\label{sec:conc_remarks}

In the second part of this two-part paper, we proposed a new algorithm for uplink link-adaptation in \gls{MU-MIMO} systems that use arbitrary receivers. This algorithm uses \gls{BMDR} as a metric for \gls{MCS} selection in \gls{MU-MIMO} systems, and it was shown to perform as well as the state-of-the art for \gls{LA} with the \gls{LMMSE} detector. More importantly, it was shown to be quite effective for \gls{LA} with the $K$-best detector which is a popular non-linear detector in the literature for which there is no previously known technique to perform full-scale \gls{LA}. We also proposed a hybrid strategy that dynamically selects the most suitable detector from a list of available detectors for improved performance. We next proposed a technique to perform \gls{PHY} abstraction in \gls{MU-MIMO} systems for arbitrary receivers. The proposed technique allows a simpler evaluation of complex non-linear receiver algorithms without significantly compromising on the accuracy of the performance metrics.

We remarked that Algorithm~\ref{alg:la} is optimal for linear detectors and is possibly optimal for the $K$-best detector. Obtaining a low-complexity algorithm that can be proven to be optimal for a general class of non-linear detectors needs further investigation. In addition, the trade-off between computational complexity and spectral efficiency of a hybrid detector obtained by solving \eqref{eq:detector_sel1} for a variety of different weights can be another interesting future research direction. Further, one might even consider exploring the usage of \gls{BMDR} in other applications such as user-pairing and \gls{MU-MIMO} scheduling, where the task is to decide which among several users can be co-scheduled together such that the throughput and latency targets for each user are satisfactorily met when a particular non-linear detector is used. 

The most computation-intensive task in both \gls{LA} and \gls{PHY} abstraction for non-linear detectors is the offline training of the \gls{BMDR}-predictor. However, this is a one-time process and the need for retraining is minimal, making the proposed technique viable for practical purposes. Non-linear detectors are playing an increasingly important role in advanced wireless technologies, especially with the advent of machine-learning-based receivers. As the uplink traffic is expected to increase significantly in the future, we expect the concepts developed in this paper to be very useful in the next-generation wireless technologies. 

\section*{Acknowledgement}
The authors would like to thank Harish Vishwanath, Suresh Kalyanasundaram, K. S. Karthik, and Chandrashekhar Thejaswi for valuable discussions on the topic, and Sivarama Venkatesan for providing a NumPy-based 3GPP channel-modeling package.